\documentclass[
reprint,
comment,
superscriptaddress,
showpacs,
amsmath,
amssymb,
aps,
prl,
floatfix,
]{revtex4-1}

\usepackage{graphicx}
\usepackage{dcolumn}
\usepackage{bm}
\usepackage{amsmath}
\usepackage{hyperref}

\usepackage[dvipsnames]{xcolor}

\begin{document}
\title{Self-Organization of Dragon Kings}

\author{Yuansheng Lin}
\email{bhlys1106@gmail.com}
\affiliation{School of Reliability and Systems Engineering, Beihang University, Beijing, China, 100191}
\affiliation{Science and Technology on Reliability and Environmental Engineering Laboratory, Beijing, China, 100191}
\affiliation{Department of Computer Science, University of California, Davis, California, USA, 95616}

\author{Keith Burghardt}
\affiliation{Department of Computer Science, University of California, Davis, California, USA, 95616}
\affiliation{Department of Political Science, University of California, Davis, California, USA, 95616}

\author{Martin Rohden}
\affiliation{Department of Computer Science, University of California, Davis, California, USA, 95616}

\author{Pierre-Andr\'e No{\"e}l}
\affiliation{Department of Computer Science, University of California, Davis, California, USA, 95616}

\author{Raissa M. D'Souza}
\affiliation{Department of Computer Science, University of California, Davis, California, USA, 95616}
\affiliation{Department of Mechanical and Aerospace Engineering, University of California, Davis, California, USA, 95616}
\affiliation{Santa Fe Institute, Santa Fe, New Mexico, USA, 87501}

\date{\today}

\begin{abstract}
The mechanisms underlying cascading failures are often modeled via the paradigm of self-organized criticality. Here we introduce a simple model where nodes self-organize to be either weak or strong to failure which captures the trade-off between degradation and reinforcement of nodes inherent in many network systems. If strong nodes cannot fail, this leads to power law distributions of failure sizes with so-called ``Black Swan" rare events. In contrast, if strong nodes fail once a sufficient fraction of their neighbors fail, this leads to ``Dragon Kings", which are massive failures caused by mechanisms distinct from smaller failures. In our model, we find that once an initial failure size is above a critical value, the Dragon King mechanism kicks in, leading to piggybacking system-wide failures. We demonstrate that the size of the initial failed weak cluster predicts the likelihood of a Dragon King event with high accuracy and we develop a simple control strategy which also reveals that a random upgrade can inadvertently make the system more vulnerable. The Dragon Kings observed are self-organized, existing throughout the parameter regime.
\end{abstract}

\maketitle
Natural and engineered systems that usually operate in a manageable regime may nonetheless be prone to rare, catastrophic events \cite{Sornette2009,Sornette2012,Carlson1999,Bak1996,Bak1987,Bak1988,Carreras2002, Hoffmann2014,wheatley2017,lorenz2009,tessone2013,RD2017}. Two categories for such events have been proposed: Black Swans, which are tail events in a power-law distribution, and Dragon Kings (DKs), which are outliers involving mechanisms absent in smaller events that occur far more frequently than a power-law would predict. The power-law distribution necessary for Black Swans to exist is often explained by self-organized criticality (SOC): a tug-of-war that poises the system close to a critical point without any need for tuning of external parameters \cite{Bak1987,Bak1988,Carreras2002,Hoffmann2014}. Although the prediction of Black Swans can sometimes beat random chance \cite{Ramos2009}, the task appears to be inherently difficult \cite{Geller1997,Taleb2005}. Despite this drawback, there are simple methods to push SOC systems away from criticality, thus reducing the size of Black Swans \cite{Cajueiro2010,Noel2013,Hoffmann2014}.

It has been proposed that DKs occur in complex systems that have low heterogeneity and strong coupling (as defined in \cite{Osorio2010}) and that, in contrast, Black Swans occur in systems with weaker coupling and higher heterogeneity. Whereas Black Swans often have no associated length- and time-scales, DK events do: there are typical places and times when DKs will and will not occur. This has been successfully applied to, for example, prediction of material failure and crashes of stock markets \cite{Sornette2009}, and has been seen in engineered systems, such as error cascades in a collection of robots \cite{Gauci2017}. Unlike Black Swans, however, it has been an open problem to control DKs in many situations and to elucidate the mechanisms underlying these often self-amplifying cascades \cite{Sornette2012}. Recent advances on controlling DKs have been based on low-dimensional models, such as coupled oscillators \cite{Cavalcante2013}, but control of DKs in models of high-dimensional complex systems has been lacking. 

In this Letter, we introduce a simple model where nodes in a network self-organize to be ``weak" or ``strong" to failure, capturing the tradeoffs between degradation and reinforcement of elements in a system. The initial failure of a random weak node can lead to a cascade of subsequent node failures. 
A weak node fails as soon as \emph{one} of its neighbors fails, and a failed weak node has small probability, $\epsilon$, to be reinforced and upgraded to a strong node upon repair.  Strong nodes independently degrade (i.e., become weak) at a slow rate. If strong nodes \emph{cannot} fail, we call the model the ``\emph{inoculation}'' (IN) model \cite{Kermack1927}. This is akin to site percolation because a failure is contained to an individual cluster of adjacent weak nodes. If strong nodes fail as soon as \emph{two} of their neighbors fail, we call the model the ``\emph{complex contagion}'' (CC) model \cite{Centola2007}. This model can lead to self-amplifying failures that cascade across clusters of weak nodes. The CC model is to our knowledge the simplest model that produces self-amplifying cascading failures, and the IN model provides a null model for baseline behavior.

We are interested in the long-term behavior: each cascade causes small changes in the number of weak and strong nodes, and both models self-organize to specific (but distinct) states. While the IN model is similar in spirit to some previous self-organized critical models of engineered systems \cite{Carreras2002, Hoffmann2014}, the CC model is expected to spontaneously generate DKs (failures of nearly the entire system) over all values of $\epsilon<1$, a prediction that we have confirmed for $\epsilon$ over several orders of magnitude.The main reason for this  difference is that the CC model enables cluster hopping cascades that occur once the first cluster of weak nodes to fail is sufficiently large, as shown herein. 
We take advantage of this finding to predict whether a small initial failure will cascade into a DK event by showing that the probability that a strong node, which bridges weak-node clusters, will have two neighbors in the initial failing cluster can be mapped onto a generalization of the birthday problem \cite{allen2014}. Once this probability is significant, then failures are likely to spread from the first weak-node cluster to subsequent weak-node clusters. More strong nodes are then likely to fail by piggybacking off of the previous failures. We can make a qualitative analogy to the gas-water phase transition in condensed matter, where droplets can nucleate. In both our model and in droplet nucleation, there is a critical size, above which the droplet or failed cluster grows almost without bound \cite{Zeng1991}, although in the CC model, clusters of any size can form (there is no analogous surface tension).

We also develop a simple targeted-reinforcement control strategy, in which we turn a few fairly well-chosen weak-nodes into strong nodes, and decrease the likelihood of DKs and other large failures by orders of magnitude. 

\paragraph{Self-organizing models.} 
The dynamics of our models depend on two competing mechanisms: degradation and reinforcement. Degradation, which represents the aging of infrastructure or an increase of load placed on them, is modeled by slowly converting strong nodes into weak ones. Conversely, reinforcement converts weak nodes that fail during a cascading event into strong nodes at rate $\epsilon$,  representing the hardening of nodes in an attempt to prevent future failures. This repair strategy mimics modern-day power grid guidelines \cite{Short2004}, where resources are allocated to places were failures happen more often. The trade-off between degradation and reinforcement drives the system to an SOC state. 

For simplicity, we consider dynamics on $3$-regular random networks with $N$ nodes, where $N$ is an even positive integer. Repeated edges and self-loops are allowed, but are rare when $N$ is large. The system size $N$ and the probability $\epsilon$ are the model's only parameters. We are particularly interested in large $N$ and small $\epsilon$, but due to motivation from real-world systems, we are also interested in finite-size effects as well as the consequences of a non-zero $\epsilon$ (i.e., having a budget for reinforcement).

\begin{figure}[t]
	\centering
  \includegraphics[width=1\columnwidth]{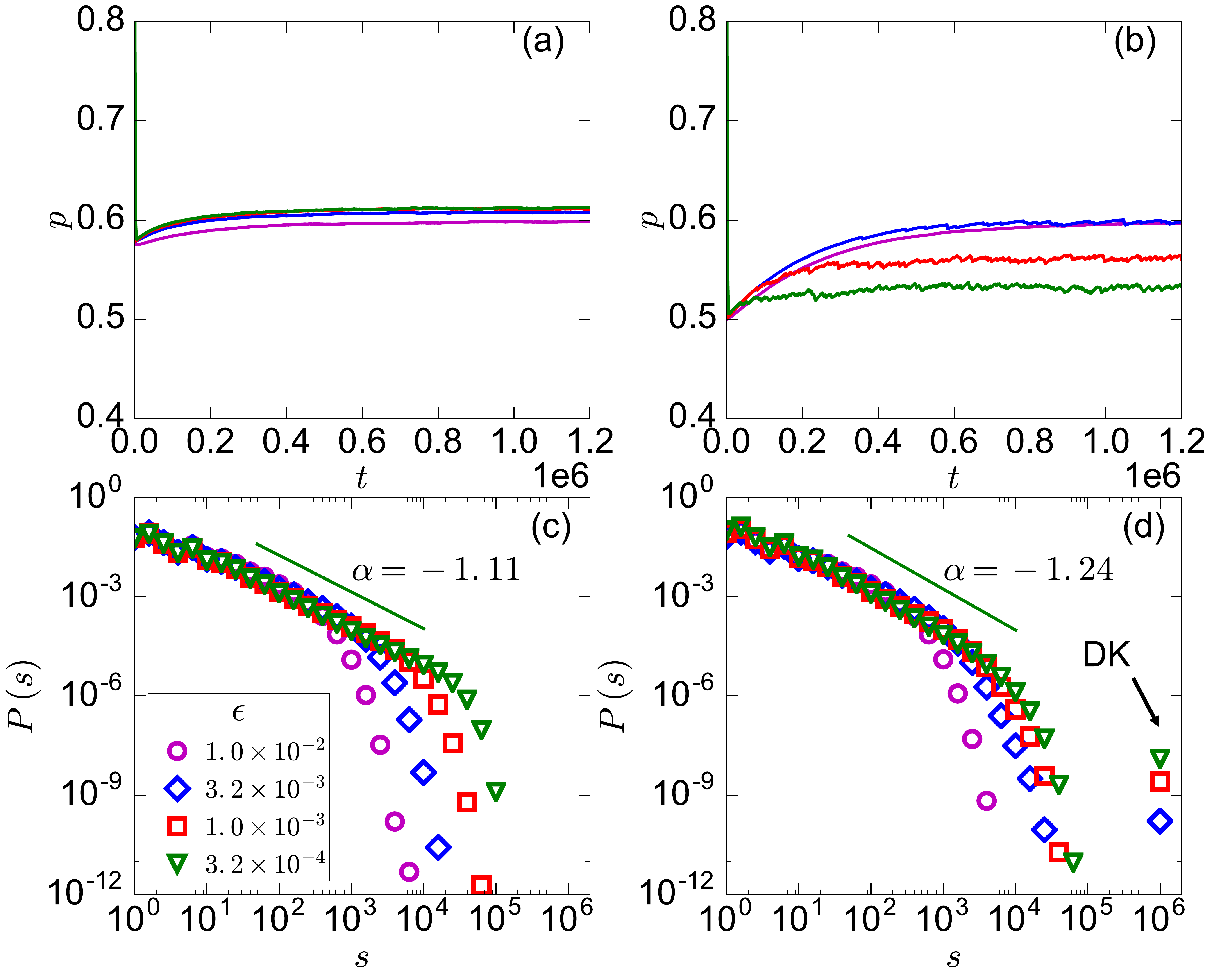}
	\caption{\label{fig:failuresize}(Color online) Self-organizing behavior and failure size. Top row: Fraction of weak nodes $p(t)$ vs. $t$ for the (a) IN and (b) CC models over individual network realizations for $N=10^6$. Bottom row: Failure size distribution for the (c) IN and (d) CC models, with the DK outliers labeled. Symbols denote results of simulations on random $3$-regular graphs averaged over ten network realizations and $15\times N$ time steps.}
\end{figure}

Both the CC and IN models follow the same general algorithm. Weak nodes fail if at least one of their neighbors fail which can cause subsequent failures. The distinction is that under the CC model, strong nodes fail if at least two of their neighbors fail, whereas in the IN model, strong nodes cannot fail. In detail, we initialize all $N$ nodes as weak and each discrete time step $0 \le t \le t_{\text{stop}}$  proceeds according to the following algorithm.
\begin{description}
  \item[Degradation] Select a node uniformly at random. If that node is strong, make it weak and proceed to the beginning of the Degradation step with $t\leftarrow t+1$. If the selected node is already weak, then it fails, and continue with the remaining three steps.
  \item[Cascade] Apply the IN or CC failure-spreading mechanism until no more failures occur. Failed nodes remain failed for the duration of the cascade. 
  \item[Repair] All failed nodes are un-failed (strong failed nodes become strong un-failed nodes, and weak failed nodes become weak un-failed nodes).
  \item[Reinforcement] Each weak node that failed at this time step has probability $\epsilon$ to become strong. Proceed back to the Degradation step with $t\leftarrow t+1$.
\end{description}

The IN model is similar to the SIRS model in epidemiology \cite{MenaLorca1992}, except that failed (i.e., infected) nodes can directly become un-failed (i.e., susceptible) again. Many other choices for initial conditions are possible, but our investigations show that the steady state behavior is independent of these choices (see SI). Because we currently initialize all nodes as weak, the sizes of the first few cascades are on the order of the system size, and numerous node upgrades take place before the system equilibrates. An important indicator that we have reached the relaxation time is the proportion of nodes that are weak at time $t$, $p(t)$, which is shown in the top row of Fig.~\ref{fig:failuresize}. We wait until well after $p(t)$ stabilizes ($5\times N$ timesteps) and then calculate failure sizes for a subsequent $15 N$ timesteps. Although we cannot prove that the model has reached equilibrium, waiting longer, and varying the initial conditions (see SI) produces quantitatively similar results. For the IN model, we find that $p(t)$ is almost independent of $\epsilon$ as $\epsilon \rightarrow 0$,  but in the CC model, the steady-state value of $p(t)$ depends on $\epsilon$.

\paragraph{Failure size distribution.} The results for the failure size distribution, $P(s)$, are illustrated in the bottom row of Fig.~\ref{fig:failuresize}, which demonstrates each model's propensity to create large events. The probability of large failures generally increases with decreasing $\epsilon$ for both the IN and CC models because, if less nodes are reinforced, cascades can more easily spread and affect larger portions of a network. For small enough $\epsilon$, we find that the cascade size distributions for the IN and CC models exhibit a power-law with exponential decay, however the CC model also has a DK tail, where \emph{over 99.9\% of nodes fail} in each DK event (cf.~Fig.~\ref{fig:failuresize}(d)). Furthermore, they appear to have two different power-law exponents: $\alpha=-1.11$ for the IN model and $\alpha=-1.24$ for the CC model when $\epsilon=3.2\times10^{-4}$ and $N=10^6$ (in comparison, traditional SOC models yield $\alpha=-1.5$ \cite{Alstrom1988,Christensen1993}).

\begin{figure}[t]
	\centering
	\includegraphics[width=1\columnwidth]{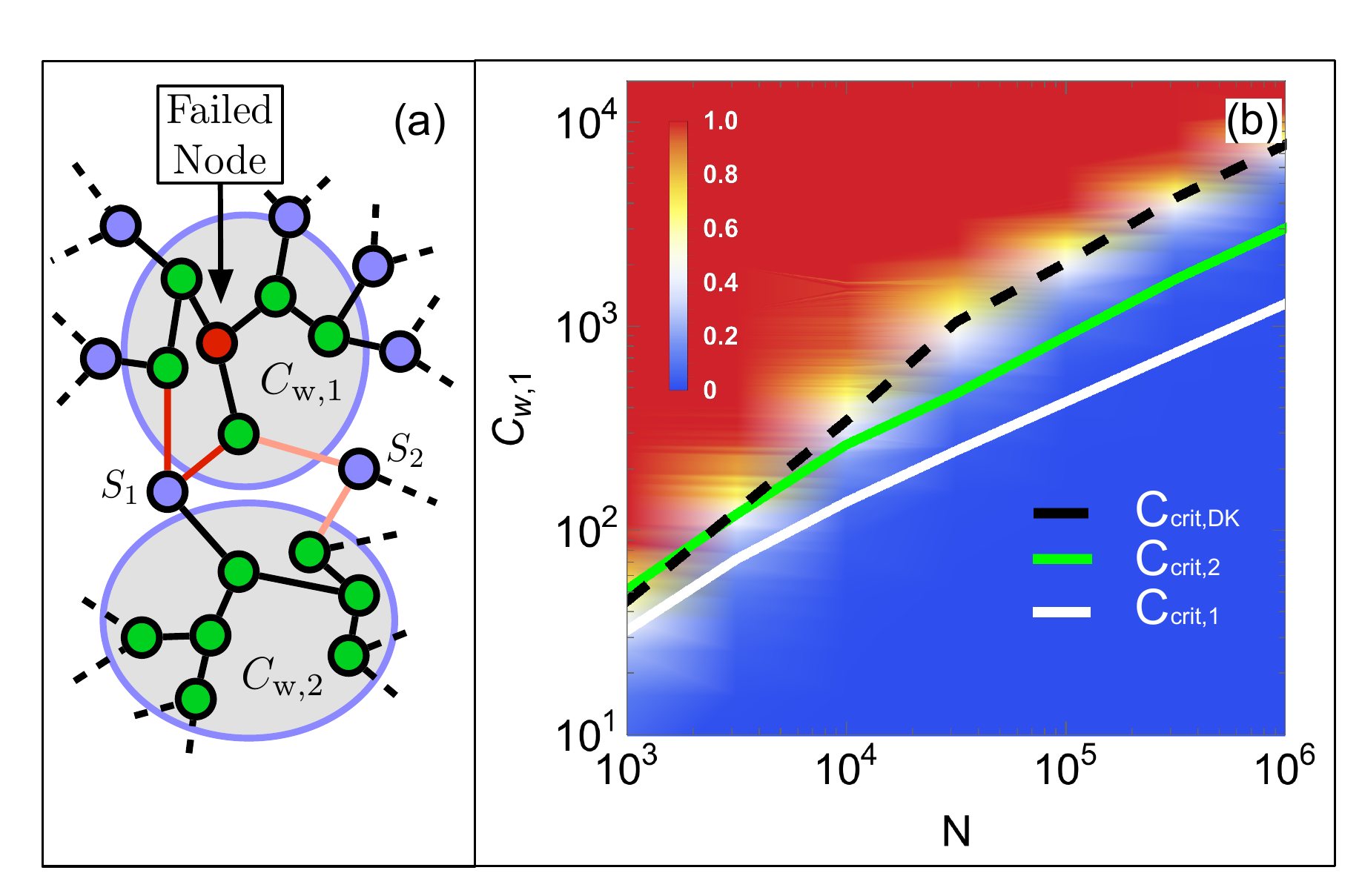}
	\caption{(Color online) DKs form by cascading failures of weak-node clusters.
(a) Weak-node clusters (circled) are surrounded by strong nodes. If one such strong node ($S_1$) has two links connecting to the same weak-node cluster, the failure of this cluster (with size $C_\text{w,1}$) will make the strong node fail, and the cascade may spread to other weak-node clusters (e.g., one of size $C_\text{w,2}$), and thus other strong nodes (e.g., $S_2$), eventually creating a DK event.
(b) A heat map of the probability a DK occurs in the CC model conditioned on $C_\text{w,1}$, the size of the weak-node cluster that first fails, versus $N$ and $C_\text{w,1}$ for $\epsilon=10^{-3}$. Black dashed line is the simulation result for $C_\text{crit,DK}$, while solid lines denote our analytic calculations of two-step failure cascades $C_\text{crit,2}$ (green line) and one-step failure cascades $C_\text{crit,1}$ (white line).}
\label{fig:PrDK}
\end{figure}

\paragraph{Dragon King Mechanism.} Why do DKs occur in the CC model? To establish a theoretical understanding of DKs, we first note that a failure in any part of a weak-node cluster makes that entire cluster fail. A necessary, but not sufficient, condition for a DK to occur is that strong nodes bridging the first failed weak-node cluster must also fail (cf.~Fig.~\ref{fig:PrDK}(a)), which we call it a one-step failure cascade. In the simplest case, only one strong node bridges two weak-node clusters. We first analyze the probability that the failure of a weak-node cluster, with size $C_\text{w,1}$, will lead to the failure of at least one bridging strong node, denoted by $S_1$, and find that $S_1$ nodes can accurately model the probability of multiple weak-node clusters failing (see SI). 

The one-step failure cascade is, however, a poor approximation of a DK event (cf.~Fig.~\ref{fig:PrDK}(b)), where cascades lead to yet more cascades (i.e., failed weak-node clusters lead to subsequent cluster failures until almost all nodes fail). To better understand DKs, we need to know whether the one-step failure cascade will lead to further failures, e.g., a two-step cascade, which requires a bridging node of type ``$S_2$'' in Fig.~\ref{fig:PrDK}(a). To obtain this probability, we first prove that the number of weak-node clusters that fail just after the first weak-node cluster fails is Poisson distributed. Furthermore, if we assume that the clusters are independent and identically distributed random variables from a scale-free distribution (which has been numerically verified from our simulation), then we can find the distribution of failed weak nodes after a one-step failure cascade, which we use to calculate the probability of two-step failure cascades, given the initial cluster size $C_\text{w,1}$. We numerically find that the critical value of $C_\text{w,1}$ such that $P(\text{two-step cascade}|C_\text{w,1}) = 1/2$, which we denote by $C_\text{crit,2}$, occurs when
\begin{equation}
C_\text{crit,2}\sim N^{0.55\pm0.01}.
\label{eqn:newtheory}
\end{equation}
Although a necessary condition, a two-step failure cascade does not always create a DK, therefore $P(\text{two-step cascade}|C_\text{w,1}) > P(\text{DK}|C_\text{w,1})$, which implies that $C_\text{crit,2}<C_\text{crit,DK}$, where $C_\text{crit,DK}$ is the critical size of $C_\text{w,1}$ such that $P(\text{DK}|C_\text{w,1}) = 1/2$. We find that these bounds agree with what we see numerically (cf.~Fig.~\ref{fig:PrDK}(b)), and $C_\text{crit,2}$ is in much closer agreement than $C_\text{crit,1}$ is to $C_\text{crit,DK}$. (Note $C_\text{crit,1}$ is defined as the critical size of $C_\text{w,1}$ such that $P(\text{one-step cascade}|C_\text{w,1}) = 1/2$). We next consider how these critical values scale with system size. Equation \eqref{eqn:newtheory} implies that $O(N^{0.55})\le C_{\text{crit,DK}} \le N$. These bounds are in agreement with the numerical scaling, in which $C_{\text{crit,DK}}\sim N^\gamma$, where $\gamma=0.59\pm0.03$ (see SI). Importantly, $C_{\text{crit,DK}}$ scales sub-linearly, therefore only a small proportion of the network needs to initially fail before a DK is likely to occur.

\begin{figure}[t]
	\centering
	\includegraphics[width=0.95\columnwidth]{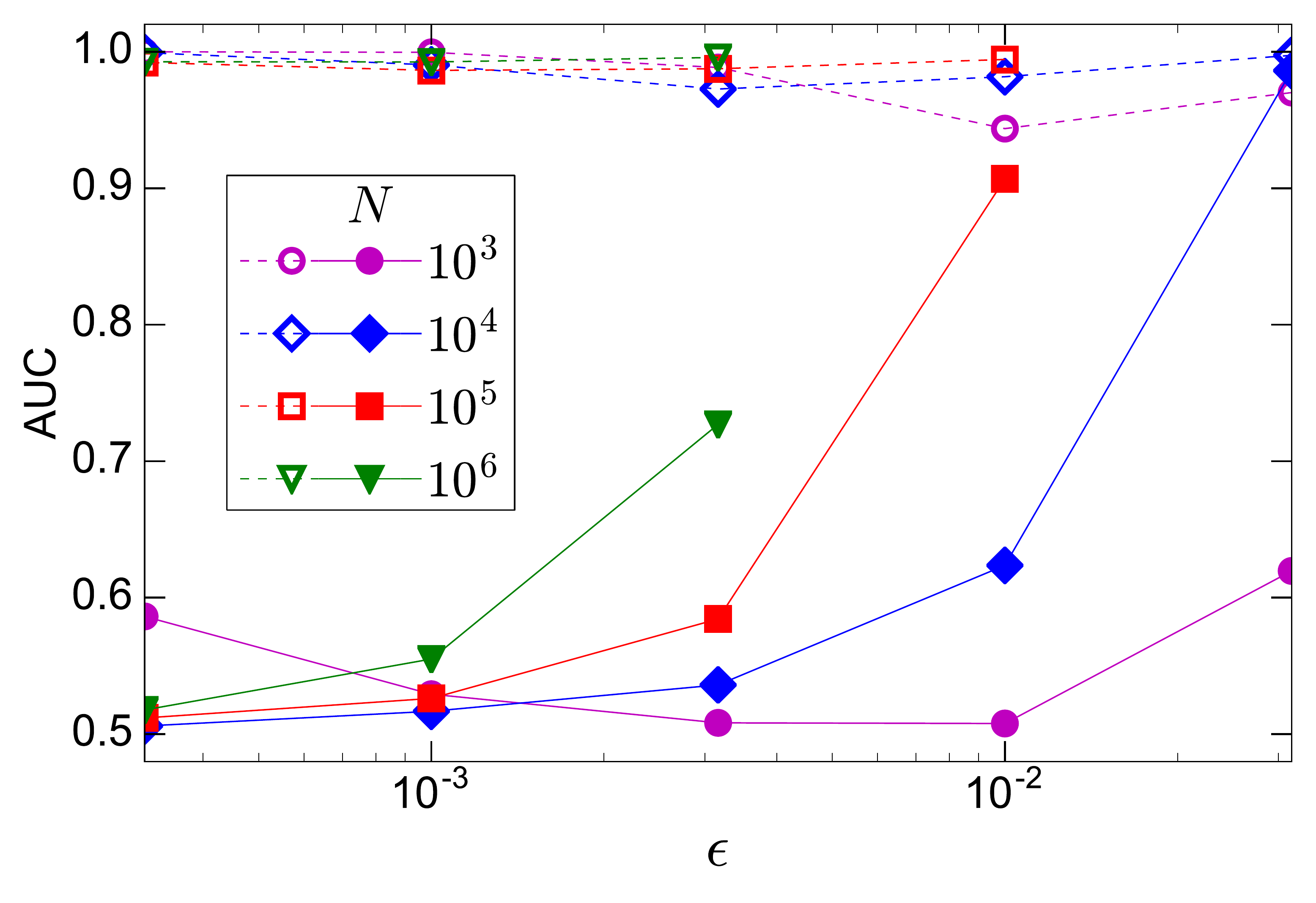}
	\caption{\label{fig:AUC}(Color online) Predicting DKs. The area under the receiver operating characteristic (AUC) for logistic models of $P(\text{DK}|p)$ (closed symbols and solid lines) and $P(\text{DK}|C_\text{w,1})$ (open symbols and dashed lines) for varying $N$ \cite{Brown2006}. 
}
\end{figure}

Finally, we can discuss how $P(\text{DK})$ varies with $N$ and $\epsilon$. First, we make simplifying assumptions that the initial failure of any weak-node cluster of size $C_\text{w,1}$ greater than $C_\text{crit,DK}$ creates a DK, and approximate $P(C_\text{w,1})$ as a power-law with an exponential cut-off, $\lambda$, that is proportional to $\epsilon$ (see SI for Fig.~S15). We find that increasing $\epsilon$ by a small amount creates an unexpectedly large percentage reduction of $P(\text{DK})$ as well as a large percentage reduction in failures that are not DKs (cf.~Fig.~\ref{fig:failuresize}). If $\epsilon$ is proportional to the cost of upgrades, then our results suggest that upgrading failed components in a system slightly more frequently can dramatically reduce the probability of large-scale failures. Surprisingly, DKs exist for any value of $\epsilon < 1$. This is because the critical weak-node cluster size increases sub-linearly with $N$, and the weak-node cluster size can be any value less than $N$, therefore there is some probability that a weak-node cluster that fails will be above the critical weak-node cluster size, which will likely trigger a DK.

When $\epsilon\rightarrow 0$ and $N$ is large, the theory suggests that $P(\text{DK})\sim N^{-0.15\pm0.02}$ (see SI), therefore DKs slowly disappear in the thermodynamic limit, but the scaling exponent is so small, that DK events are visible for almost any value of $N$ and $\epsilon$. 

\paragraph{Predicting Dragon Kings.} 
DKs are, in contrast to Black Swans \cite{Geller1997,Ramos2009}, fairly predictable \cite{Sornette2009}, although it may not be obvious what independent variables best indicate these events. For example, we find little correlation in the time between DKs (the autocorrelation is $<0.01$ for $N=10^6$, see SI), therefore, knowing the time-series of DKs will not tell is when another will necessarily occur. To answer this question, we analyze two different predictors. The first predictor is the fraction of weak nodes present in the network. The rational is that more weak nodes create larger initial failures, and therefore create more DKs. The second predictor is the size of the first weak-node cluster, $C_\text{w,1}$. We earlier established that the probability of DKs correlates with $C_\text{w,1}$, although we have yet to see whether this is adequate for predicting DKs. Both of these predictors are complimentary, because the former would tell us \emph{when} a DK might occur, while the latter would tell us \emph{where} a DK might originate.

We model $P(\text{DK}|p)$ versus $p$, and $P(\text{DK}|C_\text{w,1})$ versus $C_\text{w,1}$, respectively, using logistic regression. Unless $\epsilon$ is relatively large, $p$ is a poor predictor as based on the area under the receiver operating characteristic (AUC, cf.~Fig.~\ref{fig:AUC}) \cite{Brown2006}. Thus, predicting when a DK would occur is inherently challenging. In contrast, by knowing $C_\text{w,1}$ alone, we can predict DKs with astounding accuracy, almost independent of $N$ and $\epsilon$. The high accuracy is due to the characteristic size of the initial failure that triggers a DK, $C_\text{crit,DK}$ (see SI for figure).  This is reminiscent of previous results on controlling DKs in a system of oscillators where a trajectory straying past a particular threshold is very likely to create a DK \cite{Cavalcante2013,motter2013}. Finding the weak-node cluster size, $C_\text{w,1}$ (which is much smaller than the system size), for each weak node requires only searching locally in the network, therefore, given an initial failure, we can accurately predict whether a DK would occur with relatively little effort. Similarly, to ``tame'' DKs, we can use a simple control mechanism that requires knowing the size of just a few weak-node clusters, as seen in the next section.

\begin{figure}[t]
	\centering
	\includegraphics[width=0.92\columnwidth]{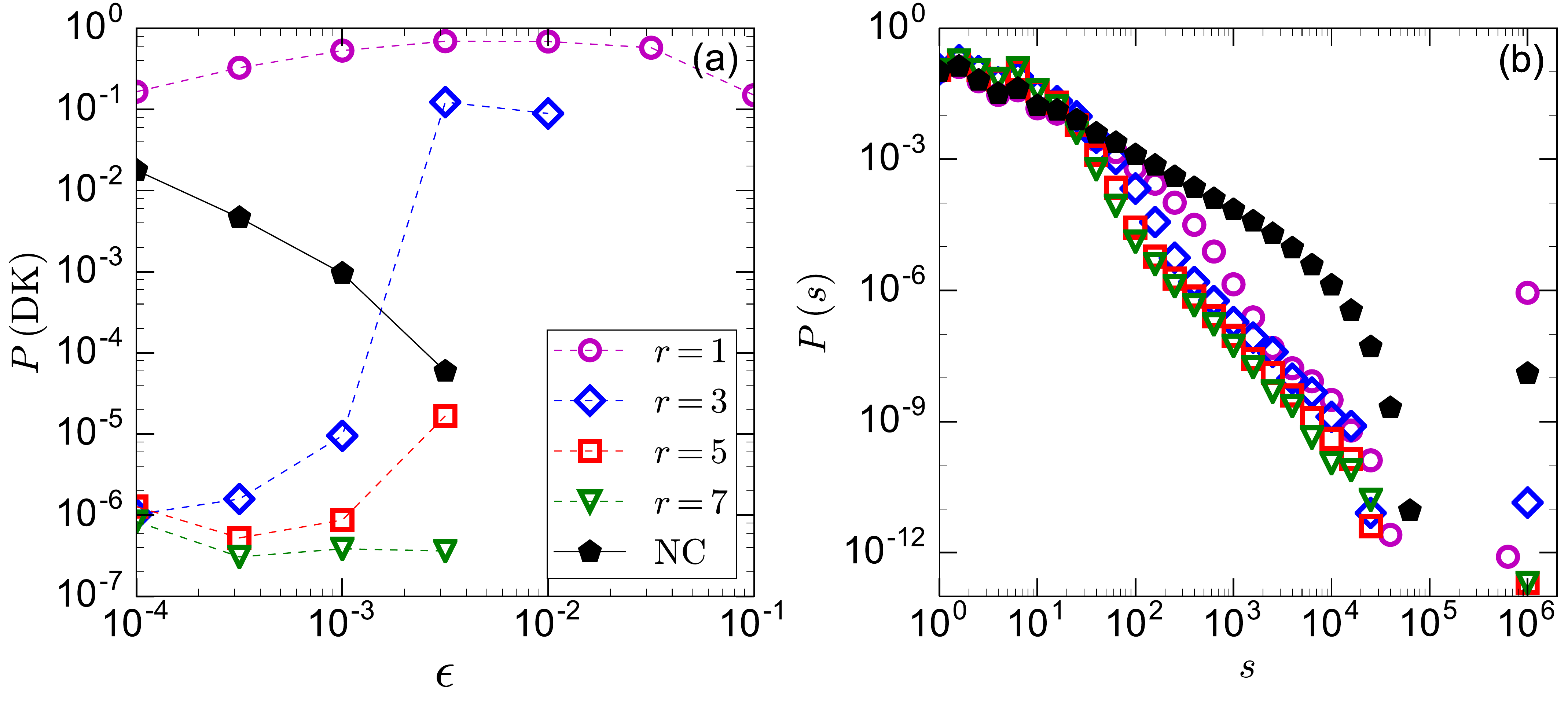}
	\caption{\label{fig:control}(Color online) Controlling DKs. (a) The probability a DK occurs over time versus $\epsilon$ in both the non-controlled scenario (NC, pentagons), and in the controlled scenario with $r$ weak-nodes chosen: $r=1$ (circles), $r=3$ (diamonds), $r=5$ (squares) and $r=7$ (triangles). Simulations are realized for $N=10^{6}$, and standard errors are smaller than marker sizes. See main text for details of the control method. (b) Failure size distributions for different control strategies and non-controlled with $\epsilon=3.2\times10^{-4}$.}
\end{figure}

\paragraph{Controlling Dragon Kings.} 
Because large weak-node cluster failures precede DKs, we can reasonably ask whether breaking up these clusters before they fail can reduce the prevalence of DKs. Assuming that the rate of node upgrades is proportional to the amount of ``money'' or effort allocated for repairing nodes, we create control strategies where this rate is kept the same on average as the non-controlled case, meaning $p(t)$ remains approximately constant. Instead of randomly reinforcing failed weak nodes, we upgrade weak nodes in large clusters by picking $r$ weak nodes and finding the size of the weak-node clusters to which they belong. The largest of these weak-node clusters is selected and with probability $1-p(t)$, a random node in that cluster is reinforced. We find that, when $r=1$, more DKs occur than without control therefore random attempts to reduce the size of failures could actually make the failures substantially worse. However, larger $r$ represents a better sampling of the cluster sizes, and a greater chance for large clusters to be broken apart, which reduces the probability of DKs by orders of magnitude (cf.~Fig.~\ref{fig:control}(a)), as well as large failures that are not DKs (cf.~Fig.~\ref{fig:control}(b)). Furthermore, the number of nodes we have to search through is only $r\times \langle C_\text{w,1}\rangle\ll N$ on average, which makes this technique applicable in systems where global knowledge of the network is lacking.

\paragraph{Discussion.} 
We have shown that DKs can self-organize in the CC model via runaway failure cascades. Moreover, this mechanism allows for DKs to be easily predicted and controlled. We believe that this model can describe a number of mechanisms, discussed below.

The CC model allows for individuals with simple contagion dynamics (weak nodes) \cite{Hodas2014}, and complex contagion dynamics (strong nodes) \cite{Centola2010}, to co-exist on a network. It assumes that agents become ``complex'' at a rate $\epsilon$ after they have adopted an idea (i.e.\ failed), which can be interpreted as agents exhibiting greater stubbornness to new ideas. The CC model suggests that agents can self-organize to a state in which global adoption (DKs) occurs surprisingly often. This could explain, for example, the mechanism of large financial drawdowns in stock markets, which are found to be DKs \cite{Sornette2009}, where social interactions, seen in stock market participation \cite{Hong2004} and foreign exchange trading \cite{Pan2012}, can convince brokers to buy or sell as a group. Some brokers will buy (sell) stock when any neighbor does, while other agents buy (sell) stock only after a sufficient fraction of their neighbors do. Our model may also represent mechanisms for cascading failures of engineered systems, where reinforcement of failed units (represented as nodes in our model) is a common practice \cite{Short2004}. 
Nodes, representing a part of a complex system, can degrade and be reinforced at slow rates to represent upgrade costs that are high and often limited to the point of making the system barely stable \cite{Carreras2002,Hoffmann2014}. 
Surprisingly, however, we find that reinforcing a system slightly more often, or selectively reinforcing nodes (the control strategy with $r>3$), creates a significant percentage drop in the frequency of DKs. In contrast, naively reinforcing nodes at random (the control strategy with $r=1$) dramatically increases the frequency of DKs.

The CC model provides a novel self-amplifying mechanism for cascading failures, and can help explain why DKs exist in the failure size distribution of real systems. Future work, however, is necessary to fully understand DKs. The research presented here provides a concrete methodology to begin studying how DKs are driven by the interplay of heterogeneity (for example the variance of node degree and the diversity of thresholds for strong nodes) and coupling (e.g., average node degree) in a principled manner, which is still in its infancy \cite{Sornette2009}. We have also not explored the effect that the mean degree, the degree distribution, or the failure threshold has on CC dynamics. Generalizing the CC model to these networks also creates additional degrees of freedom, for example the failure dynamics could depend on the minimum number of neighboring agents \cite{Centola2007}, or minimum \emph{fraction} of neighboring agents \cite{Watts2002}, that need to fail for the failure to spread to a strong node. This distinction becomes important for heavy-tailed degree distributions.

We gratefully acknowledge support from the US Army Research Office MURI award W911NF-13-1-0340 and Cooperative Agreement W911NF-09-2-0053; The Minerva Initiative award W911NF-15-1-0502; and DARPA award W911NF-17-1-0077; and financial support from the program of China Scholarships Council (No. 201506020065, Y.L.).

\section*{Supplementary Information for ``Self-Organization of Dragon Kings''}

This supplemental material provides several additional results and derivations to support the main text. In the first section, we discuss how alternative initial conditions do not appear to affect the dynamics in equilibrium. Then in the second section, we elucidate the Dragon King (DK) mechanism for the Complex Contagion (CC) model by deriving Eq.~(1) in the main text. Furthermore, we derive scaling laws to better understand whether DKs exist in the thermodynamic limit, and compare the theoretical probability of multi-step failures to simulations, conditioned on the initial failure size. In the third section, we compare the theoretical probability of DKs conditioned on $\epsilon$ and $N$ to simulations. In the fourth section, we present the failure size distributions for finite system size. In the fifth section, we show the receiver operating characteristic (ROC) curves associated with Fig.~3 in the main text. Finally, in the last section, we show that the control strategy has little effect on the fraction of weak nodes in our network, therefore, we are able to suppress DK failures while repairing the same number of nodes, on average.

\renewcommand{\thefigure}{S\arabic{figure}}
\renewcommand{\theequation}{S\arabic{equation}}
\subsection{\label{sec:I}I: Alternative Initial Conditions}

\begin{figure}[t]
	\centering
  \includegraphics[width=1\columnwidth]{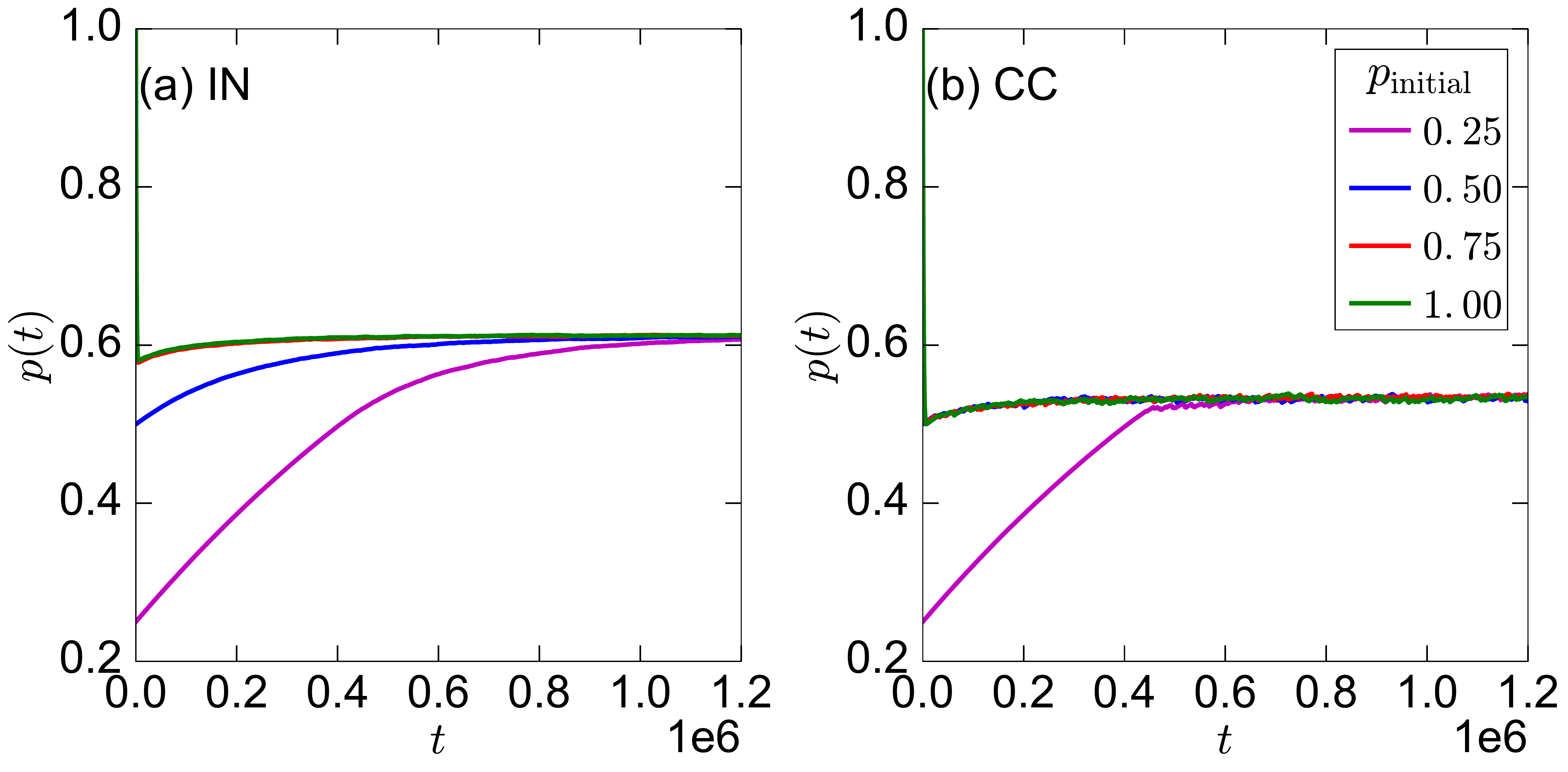}
	\caption{\label{fig:s1_pt}(Color online) Fraction of weak nodes, $p(t)$, over time for various initial conditions. We vary the initial fraction of weak nodes, $p_\text{initial}$, from $0.25$ to $1.0$ for (a) the IN model and (b) the CC model. The equilibrium value, $\langle p \rangle$, is not significantly different for various initial conditions after a time, $t$, greater than $t_{\text{relax}}=5\times N$ timesteps. In this figure, we take one network realization with $N=10^6$, $\epsilon=3.2\times 10^{-4}$.}
\end{figure} 

\begin{figure}[t]
	\centering
  \includegraphics[width=1\columnwidth]{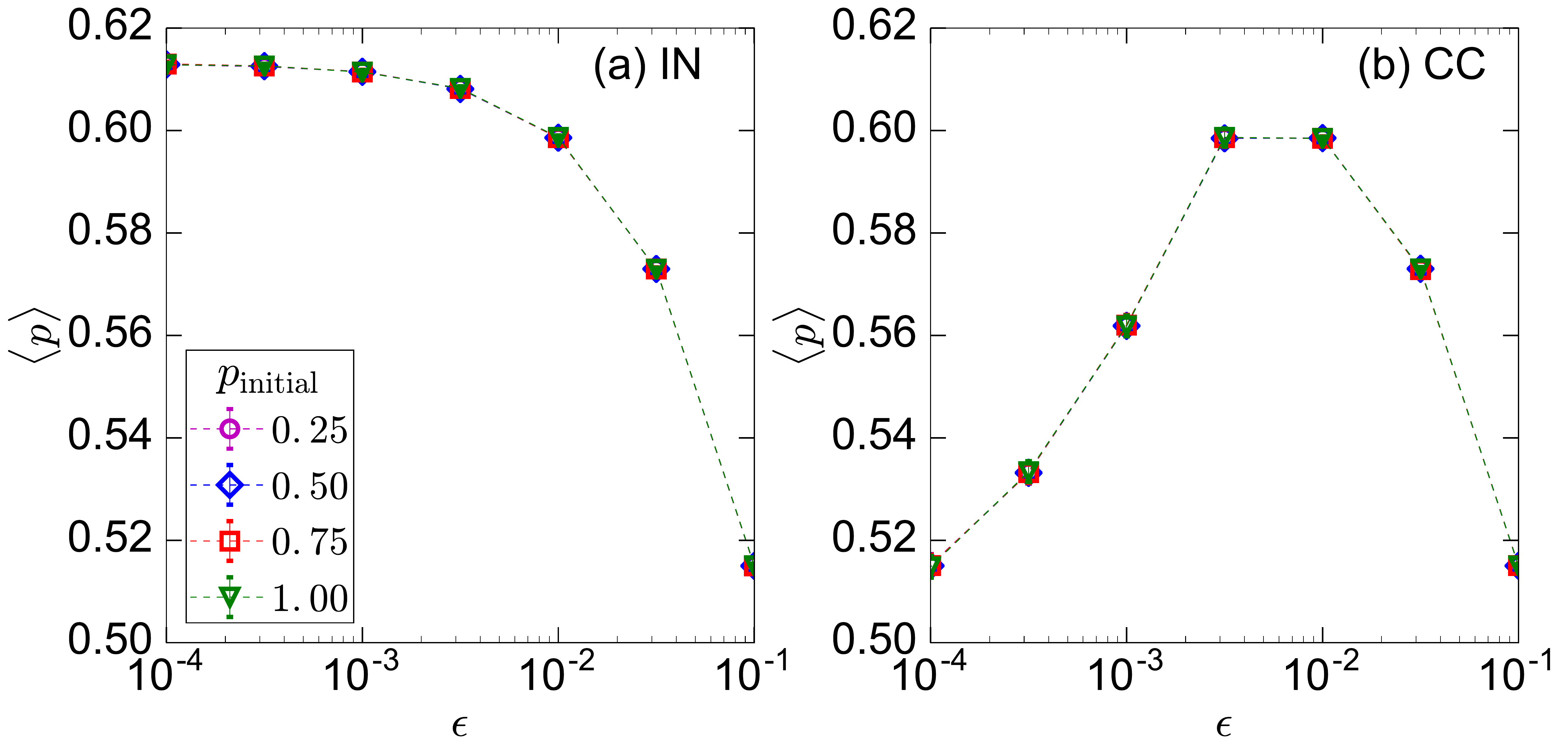}
	\caption{\label{fig:s2_paverage}(Color online) Average fraction of weak nodes after $t_{\text{relax}}=5\times N$ for different initial conditions (a) for the IN model and (b) for the CC model. The results are averaged for $N=10^6$ over $15 \times N$ timesteps and 5 network realizations.}
\end{figure}

\begin{figure}[t]
	\centering
  \includegraphics[width=0.65\columnwidth]{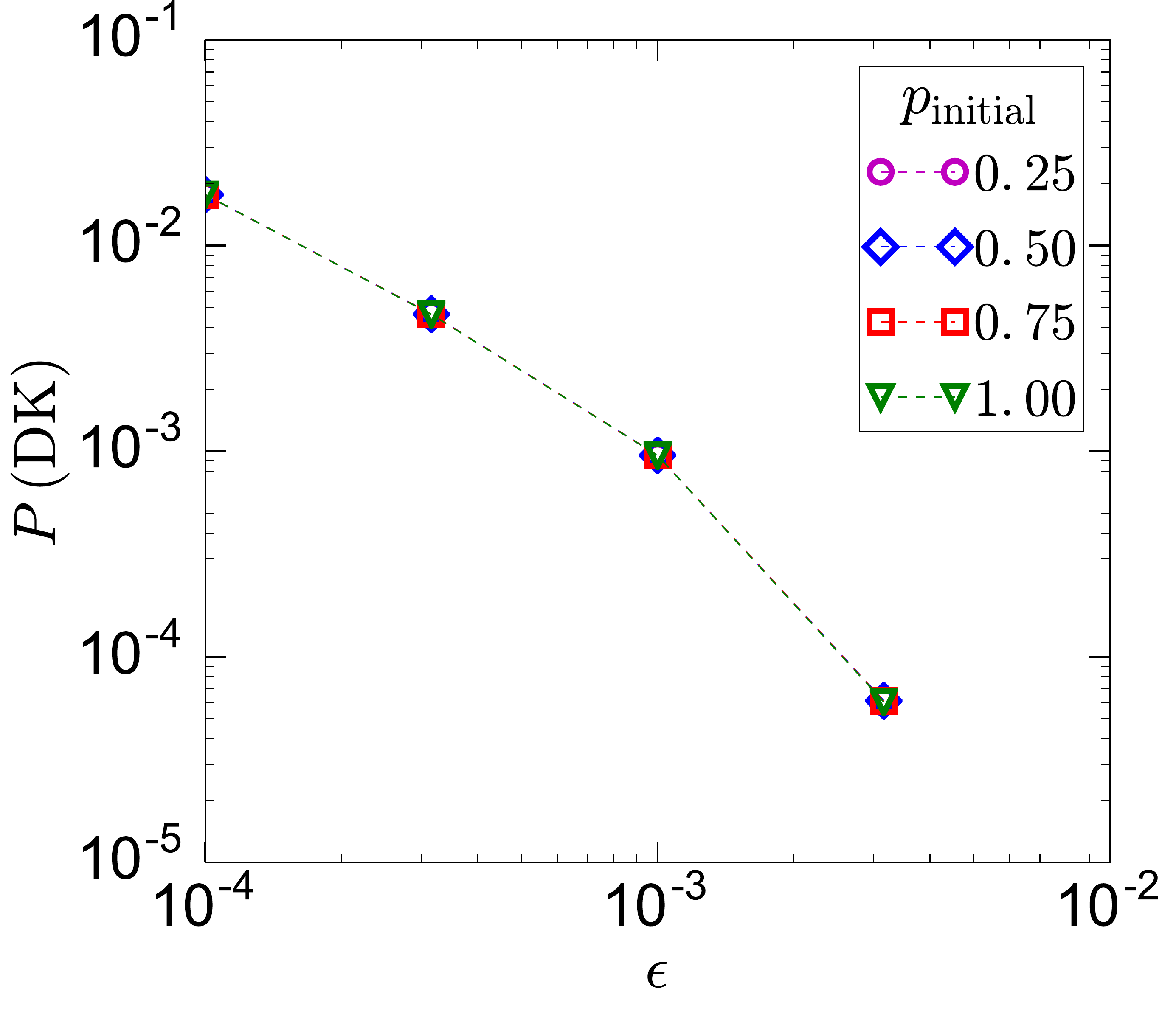}
	\caption{\label{fig:s3_DK}(Color online) Probability of DK vs. $\epsilon$ for different $p_{\text{initial}}$ in the CC model for network size with $N=10^6$. Standard errors are smaller than marker sizes. 
}
\end{figure}

We ask whether the steady state behavior of our Inoculation (IN) and CC models is independent of our initial conditions. To check this, we calculate the equilibrium fraction of weak nodes in the network, $\langle p \rangle$, and the probability of a DK, across various initial conditions as we vary $\epsilon$. In Fig.~\ref{fig:s1_pt}, we demonstrate that the fraction of weak nodes over time, $p(t)$, stabilizes to a value, $\langle p \rangle$, after a time, $t$, greater than $t_{\text{relax}}=5\times N$ timesteps across several different initial conditions in both models. To further demonstrate this, in Fig.~\ref{fig:s2_paverage} we plot the average fraction of weak nodes, $\langle p \rangle$, after $t_{\text{relax}}=5\times N$ for different reinforcement probabilities $\epsilon$ and find no statistically significant difference. Although we demonstrate that the number of weak nodes does not appear to be affected by the initial conditions, this does not guarantee that the distribution of cascade failures is unaffected. As a simple check for the CC model, Fig.~\ref{fig:s3_DK} shows that $p_{\text{initial}}$ has little effect on the probability of DK, $P(\text{DK})$. In fact, we find no statistically significant difference in $P(\text{DK})$ across initial conditions. Overall, it does not appear as though the dynamics are affected by initial conditions for $t>t_{\text{relax}}$.

\subsection{\label{sec:II}II: Dragon King Mechanism}

\subsubsection{A First Step} In this section, we present the probability of a cascade spreading from the initial weak-node cluster to any other weak-node cluster, which we call a \emph{one-step failure cascade}. We assume there are $N$ nodes, and $(1-\langle p \rangle)N$ strong nodes each of which may have between zero and three weak-node neighbors. In addition, a failure begins at a weak node within a weak-node connected cluster of size $C_\text{w,1}$. Based on simulations, the weak-node cluster is approximately a tree, therefore the number of links, $L_\text{w,1}$, from the first weak-node cluster to strong nodes should be $C_\text{w,1}+2$. Under an annealed network configuration model assumption, we sequentially connect links from the weak-node cluster to strong nodes. After $m$ links are added, the probability that a subsequent link connects to a strong node with three weak-node neighbors, is
\begin{equation}
\begin{split}
\rho_{m,j} &= \frac{3q (1-\langle p \rangle) N
-j}{(1-\langle p \rangle) N \left<k\right>-m}\\
&\approx \rho = \frac{3q}
{\left<k\right>},
\end{split}
\end{equation}
where $q$ is the average fraction of strong nodes with three weak-node neighbors, $j$ is the number of links already connected to strong nodes with three weak-node neighbors and $\left<k\right>$ is the average number of weak nodes a strong node connects to. We assume $m,j\ll (1-\langle p \rangle) N \left<k\right>$, and therefore drop second-order terms.

The probability for the $m+1^{\rm th}$ link from the weak-node cluster to connect to \emph{any} strong node that has three weak-node neighbors with one neighbor already in the same weak-node cluster is
\begin{widetext}
\begin{equation}
\begin{split}
P(\text{new $S_1$}|m) &=\sum_{j=0}^{m}\left[{m\choose j}\rho^j(1-\rho)^{m-j}\frac{2j}{(1-\langle p \rangle)N\left<k\right> - m}\right]\\
&=\frac{6 m q}{\left<k\right>}\frac{1}{(1-\langle p \rangle)N\left<k\right>-m},
\label{eqn:Pm}
\end{split}
\end{equation}
\end{widetext}
where we average over $j$. The probability that two or more links from the weak-node cluster connect to \emph{at least one} strong node with three weak-node neighbors (where we again average over all $j$) is 
\begin{equation}
\begin{split}
P(S_1|C_\text{w,1})&=1-\prod_{m=1}^{L_\text{w,1}-1} \left(1-P(\text{new $S_1$}|m)\right)\\
&\approx 1-\prod_{m=1}^{L_\text{w,1}-1} \left(1-\frac{m}{N_\text{eff}}\right)\label{eqn:birthday},
\end{split}
\end{equation}
where
\begin{equation}
N_\text{eff} = N \frac{(1-\langle p \rangle) \left<k\right>^2}{6 q}.
\label{eqn:Neff}
\end{equation}
Interestingly, the formula \eqref{eqn:birthday} is equivalent to a generalized birthday problem, where $N_\text{eff}$ is the effective number of ``days in a year'' and $L_\text{w,1}=C_\text{w,1}+2$ is the number of ``people''. Following previous literature \cite{Ahmed2000}, the critical value of $C_\text{w,1}$ is
\begin{equation}
\label{eq:Cw1}
C_\text{crit,1}=\sqrt[]{2 \text{log}(2) N_\text{eff}},
\end{equation}
which is the white line in Fig.~2 of the main text.

\subsubsection{Comparison between one-step failure cascade theory and simulations} 
\begin{figure}[t]
	\centering
  \includegraphics[width=1\columnwidth]{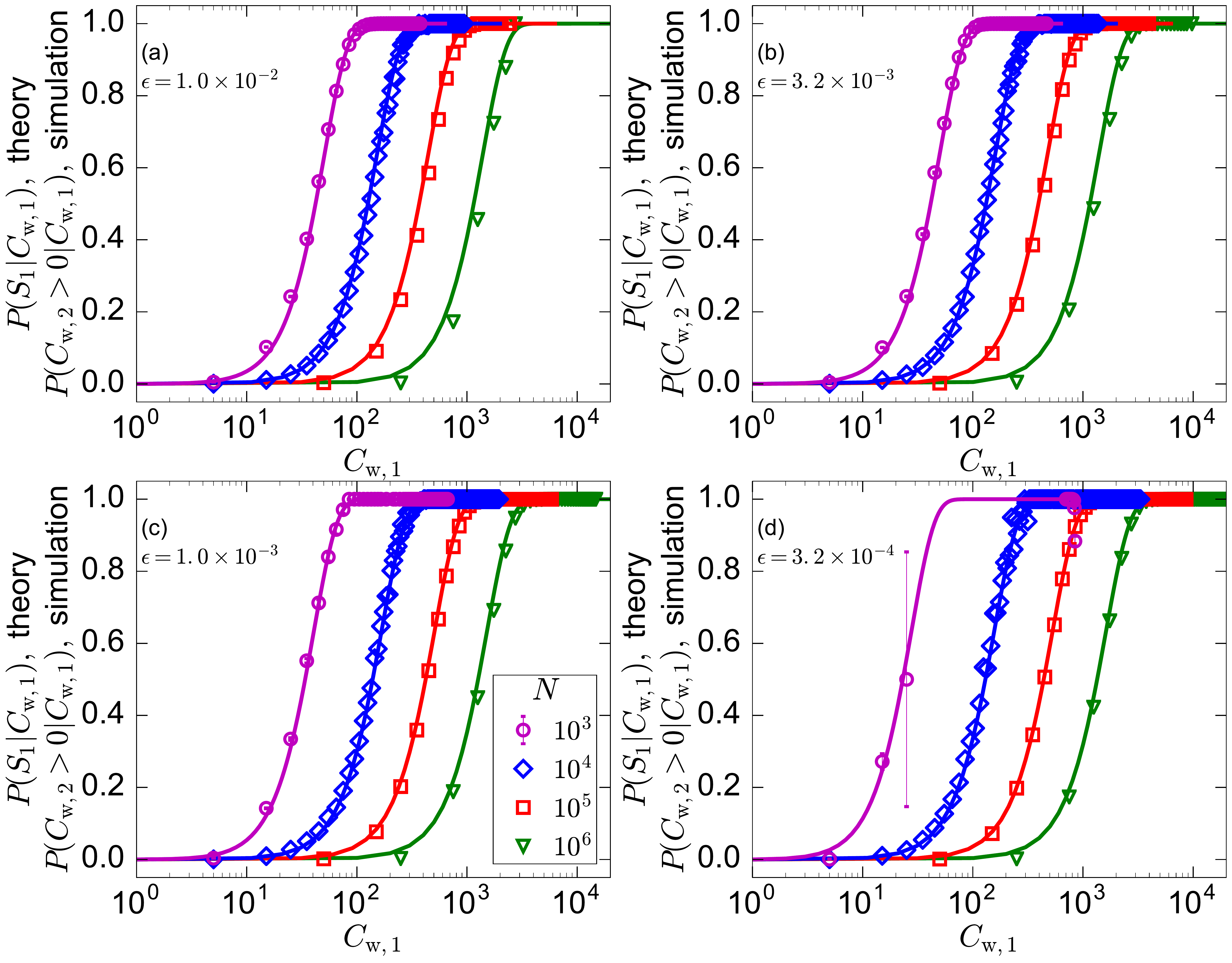}
	\caption{\label{fig:DKcompare}(Color online) The probability of failure spreading out from the initial failed weak-node cluster. Open markers represent the probability that a failure spreads from the first weak-node cluster to any other weak-node clusters, based on model simulations. Solid lines represent theoretical results of $P(S_1|C_\text{w,1})$, which is the simplest way a failure can spread between weak-node clusters. Standard errors are smaller than marker sizes, except for $N=10^3$ and $\epsilon=1.0\times10^{-3}$. 
}
\end{figure}

Next, we compare the one-step failure cascade theory (Eq.~\eqref{eqn:birthday}) to one-step failure cascade simulations (Fig.~\ref{fig:DKcompare}). We find that theory and simulation results match well.

\begin{figure}[tb]
	\centering
  \includegraphics[width=1\columnwidth]{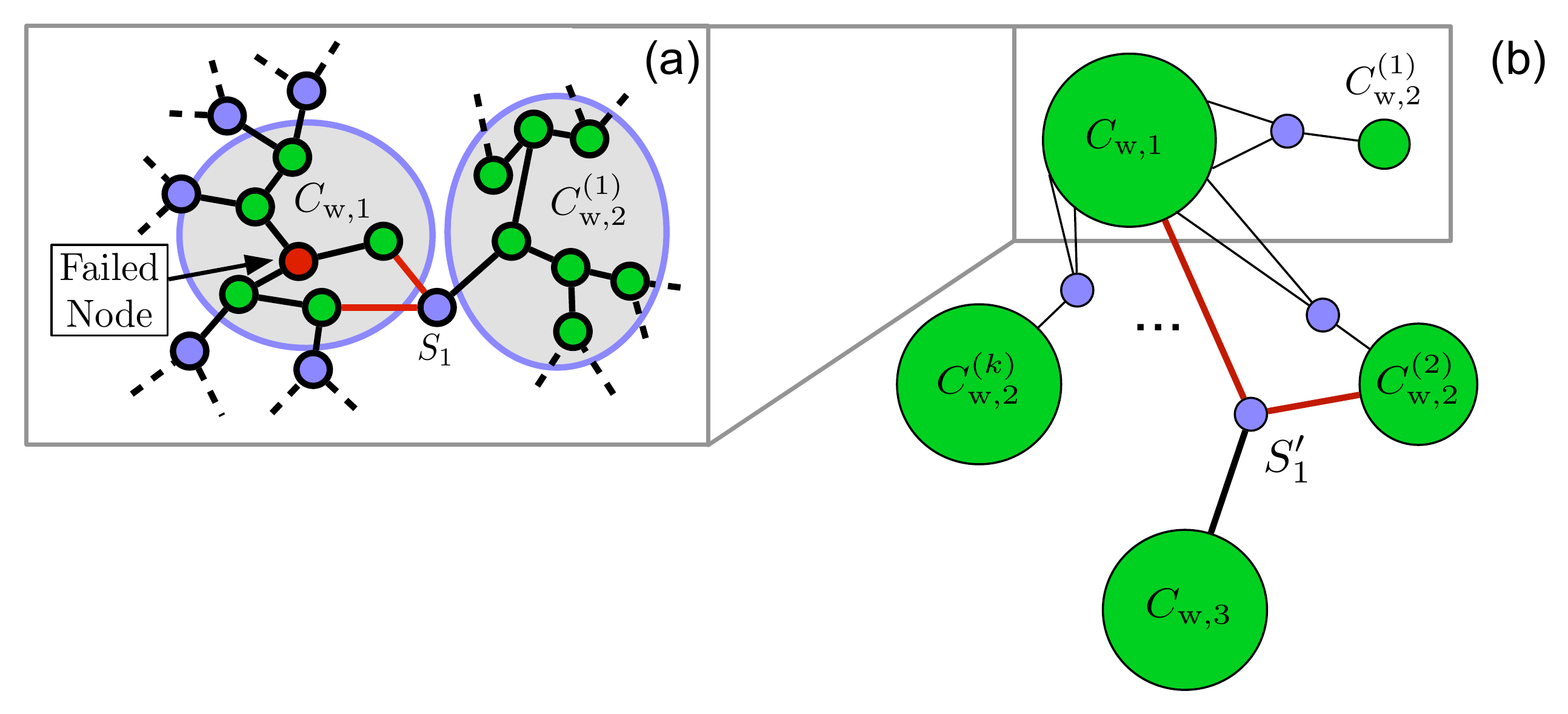}
	\caption{\label{fig:Cw3Schematic}(Color online) How cascades occur. (a) Failures spread from a single weak-node to an entire connected cluster of weak-nodes. Strong nodes are usually resistant to failure, unless they connect to the same failed cluster ($S_1$). (b) Once enough weak-node clusters fail, there is a chance that failures can spread to yet more clusters (of size $C_\text{w,3}$) via strong nodes that bridge failed weak clusters ($S_1'$).
}
\end{figure}

\subsubsection{ Going One Step Further...}

It is important to not only understand how the first step in a failure cascade occurs, but also how the additional steps of a failure cascade which might ultimately results in a DK event. As a next step, we ask what is the probability for a \emph{two-step failure cascade}, in which a weak-node cluster fails after a one-step failure cascade. The simplest way this can occur is when a strong node with three weak-node neighbors fails (see $S_1'$ in Fig.~\ref{fig:Cw3Schematic}(b)), because it can bridge the failed cluster and a new weak-node cluster. In order to find this probability, we will need to find the probability that the $S_1'$ node will connect to two failed nodes. First, to determine how many nodes failed, we first recall that the probability for the first strong nodes to fail in the cascade, $S_1$ in Fig.~\ref{fig:Cw3Schematic}(a), is a generalized birthday problem (see previous subsections), therefore the probability for $k$ nodes like $S_1$ to fail is Poisson distributed, for large $N_\text{eff}$ and $L_\text{w,1}$:
\begin{equation}
\label{eq:Cw2dist}
P_{S_1}(k) = e^{-\lambda_\text{w,1}} \frac{\lambda_\text{w,1}^k}{k!},
\end{equation}
where
\begin{equation}
\lambda_\text{w,1}=\frac{L_\text{w,1} (L_\text{w,1}-1)}{2 N_\text{eff}}.
\end{equation}
Intuitively, this is because each event (the failure of $S_1$-like nodes) is statistically independent, and occurs with a low probability ($P(C_\text{w,1}>C_\text{crit,DK})$ is small), and there are lots of opportunities for the event to occur ($N_\text{eff}$ is large). For each weak-node cluster, $i$, to fail from an $S_1$-like strong node, let the size of the cluster be $C_\text{w,2}^{(i)}$, and let $C_\text{w,2} = \sum_{i=1}^k C_\text{w,2}^{(i)}$, then the probability $C_\text{w,2} >0$ is
\begin{equation}
P(S_1|C_\text{w,1}) = 1-e^{-\lambda_\text{w,1}}. \label{eqn:birthdayapprox}
\end{equation}
This can also be derived from Eq.~\eqref{eqn:birthday}, by taking
\begin{equation}
\begin{split}
\prod_{m=1}^{L_\text{w,1}-1} \left(1-\frac{m}{N_\text{eff}}\right)
&=e^{\text{log}\left(\prod_{m=1}^{L_\text{w,1}-1} \left(1-\frac{m}{N_\text{eff}}\right)\right)}\\
&=e^{\sum_{m=1}^{L_\text{w,1}-1}\text{log} \left(1-\frac{m}{N_\text{eff}}\right)}\\
&\approx e^{-L_\text{w,1}(L_\text{w,1}-1)/(2 N_\text{eff})}.
\end{split}
\end{equation}
Using this approximation, we can directly solve for when $P(S_1|C_\text{w,1})=1/2$ (Eq.~\eqref{eq:Cw1}).

\begin{figure}[t]
	\centering
  \includegraphics[width=1\columnwidth]{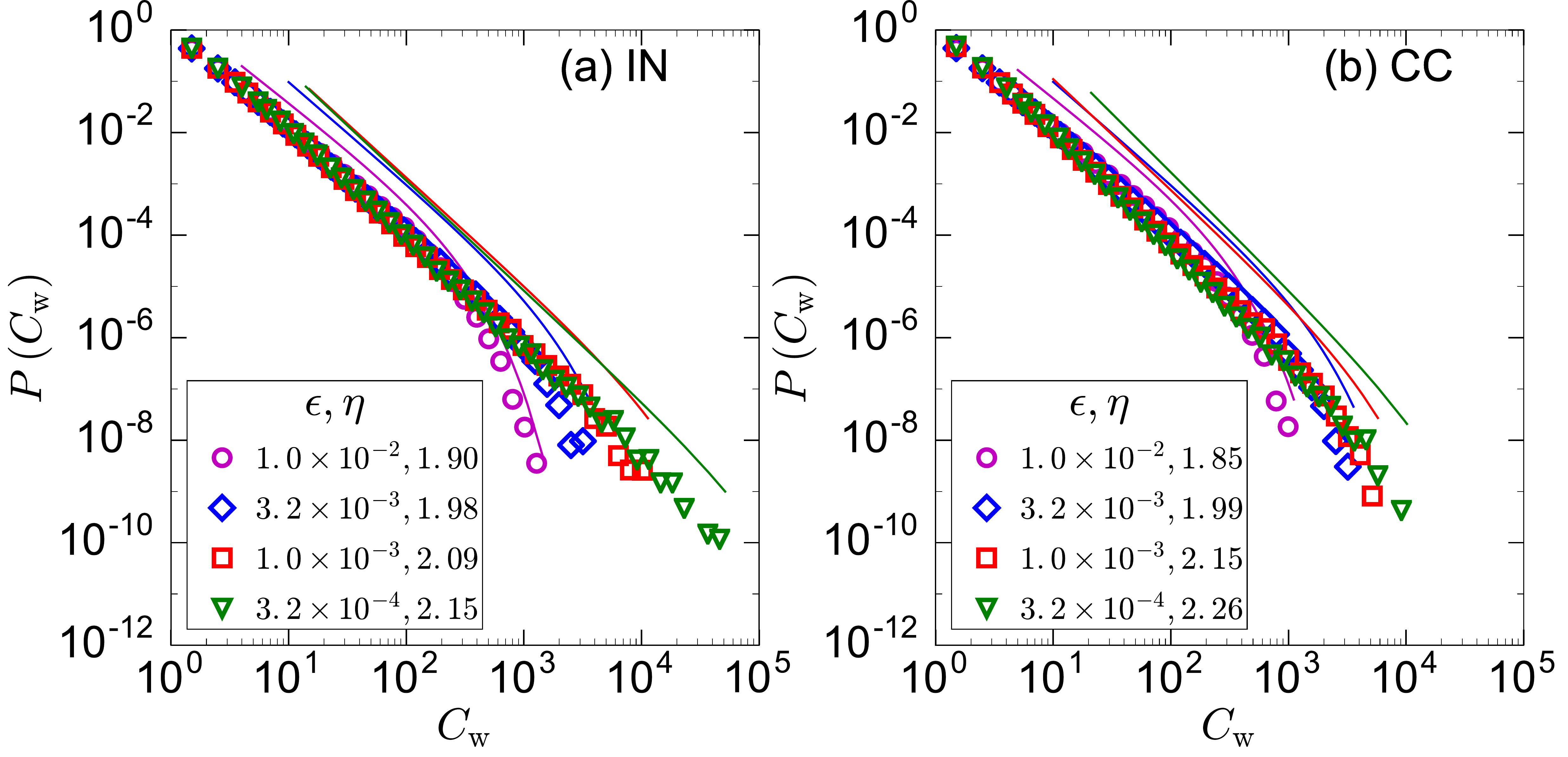}
	\caption{\label{fig:weak_cluster}(Color online) Weak cluster size distribution, $P(C_\text{w})$, for $N=10^6$ and 10 network realizations at $t=9\times N$. The parameter $\eta$ is fitting exponent of the maximum likelihood power-law with exponential tail. 
}
\end{figure}

Now that we know the distribution of weak-node clusters that first fail (Eq.~\eqref{eq:Cw2dist}), we need to know how many nodes are in each weak-node secondary cluster fail. For each cluster of size $C_\text{w,2}^{(i)}$, there are $L_\text{w,2}^{(i)} = C_\text{w,2}^{(i)}+2$ edges connected to strong nodes. Furthermore, we find empirically that the size distribution of weak-node clusters is
\begin{equation}
\label{eqn:NaiveSizeDist}
P(C_\text{w}) \sim C_\text{w}^{-\eta},
\end{equation}
where $\eta = 2.26\pm 0.03$ (measured for $\epsilon=3.2\times 10^{-4}$ and $N=10^6$, see Fig.~\ref{fig:weak_cluster}). Note, however, that the size distribution of $C_\text{w,2}^{(i)}$ is not Eq.~\eqref{eqn:NaiveSizeDist}, because the number of opportunities to connect to a node of size $C_\text{w,2}^{(i)}$ is proportional to the number of links emanating from the cluster, which increases as $L_\text{w,2}^{(i)}$. Namely, $S_1$ is a strong node with three weak-node neighbors, and there are on average  $q L_\text{w,2}^{(i)}$ strong nodes with three weak-node neighbors connected to a cluster of size $C_\text{w,2}^{(i)}$. Therefore, the probability to connect to \emph{any} cluster of size $C_\text{w,2}^{(i)}$, $P_\text{w,2}(C_\text{w,2}^{(i)})$, is
\begin{equation}
\begin{split}
\label{eq:Pw2}
P_\text{w,2}(C_\text{w,2}^{(i)}) &= \frac{P(C_\text{w} = C_\text{w,2}^{(i)}) q L_\text{w,2}^{(i)}}{\langle q L_\text{w,2}^{(i)}\rangle}\\
&= \frac{P(C_\text{w,2}^{(i)}) L_\text{w,2}^{(i)}}{\langle L_\text{w,2}^{(i)}\rangle}.
\end{split}
\end{equation}

This is intuitively similar to ``excess degree'' seen in random network literature (Eq.~(22) in \cite{Newman2003}).

\begin{figure}[tb]
	\centering
  \includegraphics[width=1\columnwidth]{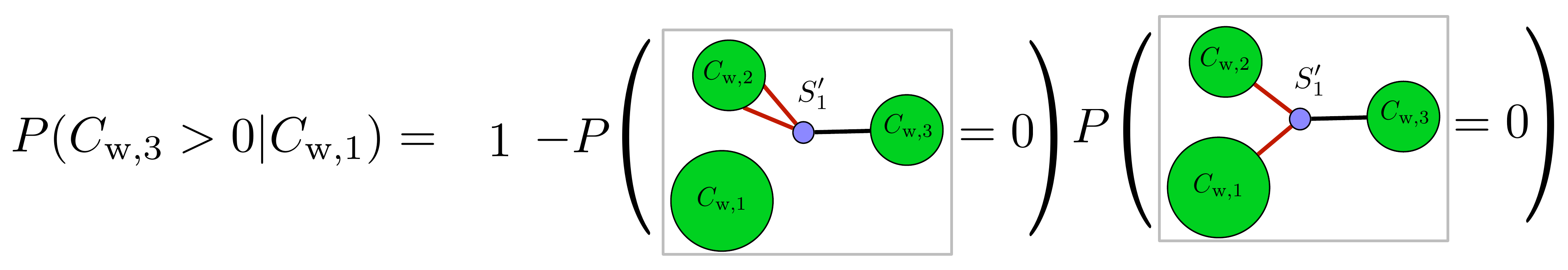}
	\caption{\label{fig:PCw3}(Color online) The probability of a two-step failure cascade. %Because all $S_1$ nodes connected to the initial weak-node cluster of size $C_\text{w,1}$ are used to fail subsequent clusters of size $C_\text{w,2} = \sum_{i=1}^k C_\text{w,2}^{(i)}$, indirect failures ($C_\text{w,3}>0$) can only occur via $S_1'$ nodes that (1) span from the initial failed weak-node cluster to the entire clusters with total size $C_\text{w,2}$ or (2)  between nodes in the clusters with size $C_\text{w,2}$. This probability is therefore $1-$ (probability that condition (1) does not occur) $\times$ (probability that condition (2) does not occur).
}
\end{figure}

$S_1'$-like nodes can only appear through one of two conditions: (1) an $S_1'$ node is created from two links in the non-initial clusters (left probability in Fig.~\ref{fig:PCw3}), or (2) an $S_1'$ node spans the initial weak-node cluster, and a newer cluster (right probability in Fig.~\ref{fig:PCw3}). Therefore $P(C_\text{w,3}>0|C_\text{w,1})$ is $1$ minus the probability that both conditions do not occur.

Condition (1) is the simpler of the two to calculate, because the derivation for the equation is very similar to that of equation \eqref{eqn:birthdayapprox}. For each secondary cluster, the number of links is $L_\text{w,2}^{'(i)} = C_\text{w,2}^{(i)}+2 - 1$, where we subtract one because one link is used to connect to the initial cluster, therefore $L_\text{w,2}' = \sum_{i=1}^k L_\text{w,2}^{'(i)} =  \sum_{i=1}^k C_\text{w,2}^{(i)}+ k$, and 
\begin{equation}
P(C_\text{w,3}=0|\text{condition }(1)) = e^{-\lambda_\text{w,2}},
\end{equation}
where 
\begin{equation}
\lambda_\text{w,2} = \frac{L_\text{w,2}' (L_\text{w,2}'-1)}{2 N_\text{eff}}.
\end{equation}

To solve for condition (2), note that there are $L_\text{w,1}' = C_\text{w,1}+2 - 2 k = C_\text{w,1}- 2 (k-1)$ links available from the initial cluster, because $2 k$ links were used to connect to $S_1$-like nodes. There are $q L_\text{w,1}'$ links from the initial weak-node cluster to strong nodes with three weak-node neighbors, therefore, there are $2 q L_\text{w,1}'$ opportunities for strong nodes with three weak-node neighbors to connect to any secondary failed weak-node clusters. The probability for one link from the initial cluster to connect to any of the secondary failed weak-node clusters is 
\begin{equation}
P(1~\text{link}\rightarrow C_\text{w,2}) = \frac{q L_\text{w,2}'}{q p N},
\end{equation}
where the denominator is the total number of links from weak-node clusters to strong nodes with three weak-node neighbors. This implies that the probability at least one strong node has a link in the initial and secondary clusters is
\begin{equation}
P(C_\text{w,3}>0|C_\text{w,1}\leftrightarrow C_\text{w,2}) = 1 - \left(1-\frac{q L_\text{w,2}'}{q p N}\right)^{2 q L_\text{w,1}'}.
\end{equation}

Similar logic from the perspective of the secondary weak-node clusters suggests that
\begin{equation}
P(C_\text{w,3}>0|C_\text{w,2}\leftrightarrow C_\text{w,1}) = 1 - \left(1-\frac{L_\text{w,1}'}{p N}\right)^{2 q L_\text{w,2}'}.
\end{equation}

We therefore have a paradox. We expect that the probability for a strong node to span initial and secondary clusters should be independent of the order we choose to connect them (the probability of the secondary clusters connecting to the initial cluster should be the same as the initial to the secondary clusters). However, if we make an ansatz that $1\ll L_\text{w,1}'\ll N$ and $1\ll L_\text{w,2}'\ll N$,
we can take the Taylor series of either equation and approximate the sum as
\begin{equation}
\begin{split}
P(C_\text{w,3}>0|C_\text{w,1}\leftrightarrow C_\text{w,2}) &\approx P(C_\text{w,3}>0|C_\text{w,2}\leftrightarrow C_\text{w,1})\\
&\approx 1 - e^{-\lambda_\text{w,1$\leftrightarrow$2}},
\end{split}
\end{equation}
or
\begin{equation}
P(C_\text{w,3}=0|\text{condition }(2)) \approx e^{-\lambda_\text{w,1$\leftrightarrow$2}},
\end{equation}
where 
\begin{equation}
\lambda_\text{w,1$\leftrightarrow$2} = \frac{2 q L_\text{w,1}'L_\text{w,2}'}{p N}.
\end{equation}
This equation is also order-independent, as we expect. We can therefore write $P(C_\text{w,3}>0|C_\text{w,1})$ as the probability that neither condition (1) nor condition (2) occurs (i.e., Fig.~\ref{fig:PCw3}). Recall that this is the probability that at least one $S_1'$-like node occurs over all values of $C_\text{w,2}^{(1)}$, $C_\text{w,2}^{(2)}$, ..., $C_\text{w,2}^{(k)}$. The probability of $k$ different $S_1$-like nodes is $P_{S_1}(k)$, while $P_\text{w,2}(C_\text{w,2}^{(1)},C_\text{w,2}^{(2)},...,C_\text{w,2}^{(k)}) = P_\text{w,2}(C_\text{w,2}^{(1)})\times P_\text{w,2}(C_\text{w,2}^{(1)})\times ...\times P_\text{w,2}(C_\text{w,2}^{(k)})$, where $P_\text{w,2}(C_\text{w,2}^{(i)})$ is Eq.~\eqref{eq:Pw2}. Therefore, if we approximate $C_\text{w,2}^{(i)}$ as a continuous variable, $P(C_\text{w,3}>0|C_\text{w,1})$ can be written as
\begin{widetext}
\begin{equation}
\begin{split}
\label{eq:simplePrw3}
&P(C_\text{w,3}>0|C_\text{w,1}) = \\
&\sum_{k=1}^\infty P_{S_1}(k) \int_{C_\text{w,2}^{(1)}=1}^\infty \int_{C_\text{w,2}^{(2)}=1}^\infty ... \int_{C_\text{w,2}^{(k)}=1}^\infty  P(C_\text{w,3}>0|C_\text{w,1},C_\text{w,2}^{(1)},...,C_\text{w,2}^{(k)})  P_\text{w,2}(C_\text{w,2}^{(1)})\times P_\text{w,2}(C_\text{w,2}^{(1)})\times ...\times P_\text{w,2}(C_\text{w,2}^{(k)}),
\end{split}
\end{equation}
where
\begin{equation}
P(C_\text{w,3}>0|C_\text{w,1},C_\text{w,2}^{(1)},...,C_\text{w,2}^{(k)}) = 1 - e^{-\lambda_\text{w,2}-\lambda_\text{w,1$\leftrightarrow$2}}.
\end{equation}
In Eq.~\eqref{eq:simplePrw3}, we notice that the lower bound of each integral is $1$ simply because $C_\text{w,2}^{(i)} \ge 1$, i.e., there must be at least one node. Using the above findings, Eq.~\eqref{eq:simplePrw3} can be written more compactly as 
\begin{equation}
P(C_\text{w,3}>0|C_\text{w,1}) =  \sum_{k=1}^\infty P_{S_1}(k)\left(1 - \int_{C_\text{w,2}^{(1)},C_\text{w,2}^{(2)},...C_\text{w,2}^{(k)}=1}^\infty  e^{-\lambda_\text{w,2}-\lambda_\text{w,1$\leftrightarrow$2}} \prod_{i=1}^{k} P_\text{w,2}(C_\text{w,2}^{(i)}) dC_\text{w,2}^{(i)} \right).
\end{equation}
\end{widetext}
As a sanity check, if $\lambda_\text{w,2} = \lambda_\text{w,1$\leftrightarrow$2} \rightarrow \infty $, which is the unrealistic condition that $C_\text{w,3}>0$ whenever $C_\text{w,2}>0$, the equation reduces to Eq.~\eqref{eqn:birthdayapprox}. In this equation, we recall that $C_\text{w,2}^{(i)}$ are all independent, and $\lambda_\text{w,2}$ and $\lambda_\text{w,1$\leftrightarrow$2}$ are defined to be a function of $\sum_{i=1}^k C_\text{w,2}^{(i)}$. Sadly, however, this integral is not analytically tractable, therefore we numerically determine the integral using Mathematica. For large $k$, we suffer from the curse of dimensionality, therefore we use a cutoff: we ignore any $k$ where $P_{S_1}(k)<\delta$, where $\delta = 10^{-4}$.

\subsubsection{Comparison of two-step failure cascade theory and simulations}

\begin{figure}[tb]
	\centering
  \includegraphics[width=1\columnwidth]{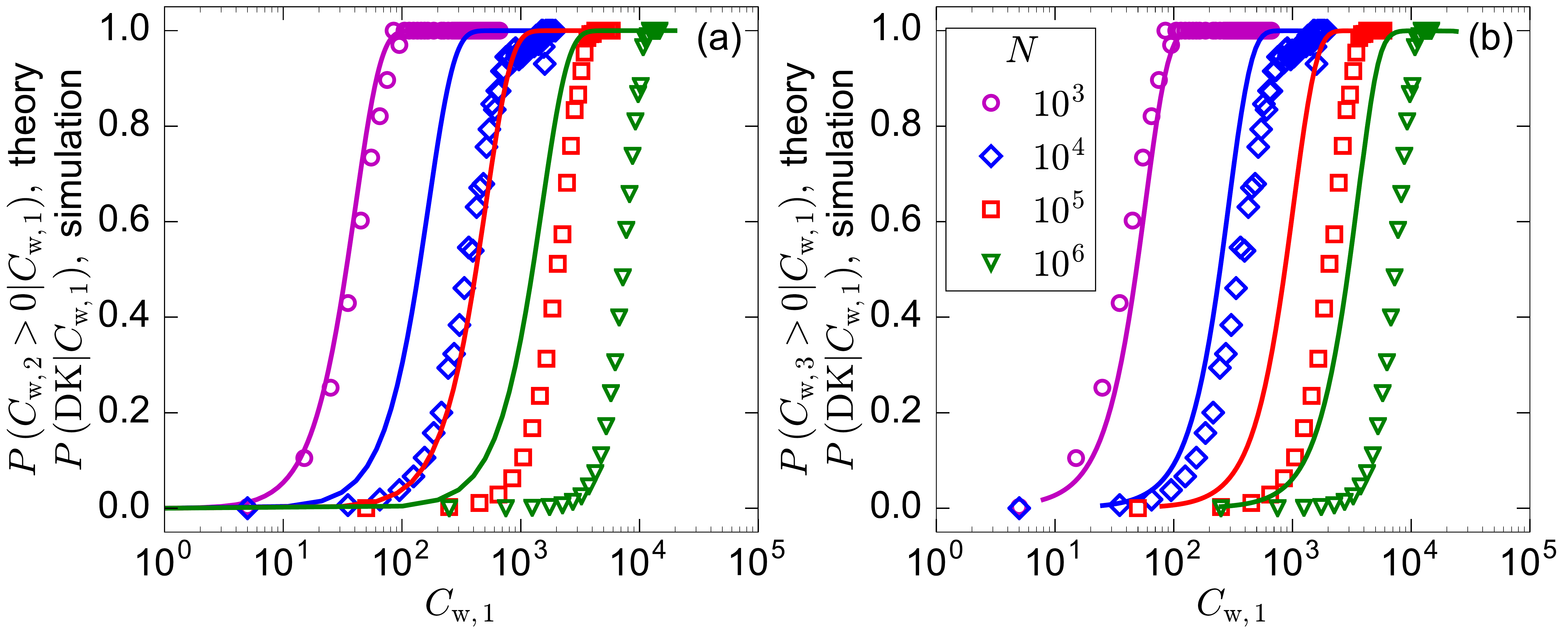}
	\caption{\label{fig:PCw2Cw3VsCw1}(Color online) Probability of one-step and two-step failure cascades and DKs versus $C_\text{w,1}$. Solid lines are the theoretical probabilities of (a) one-step and (b) two-step failure cascades, and plot markers are the simulation-based probabilities of DKs versus $C_\text{w,1}$ for $\epsilon=1.00\times10^{-3}$.
}
\end{figure}

\begin{figure}[t]
	\centering
  \includegraphics[width=1\columnwidth]{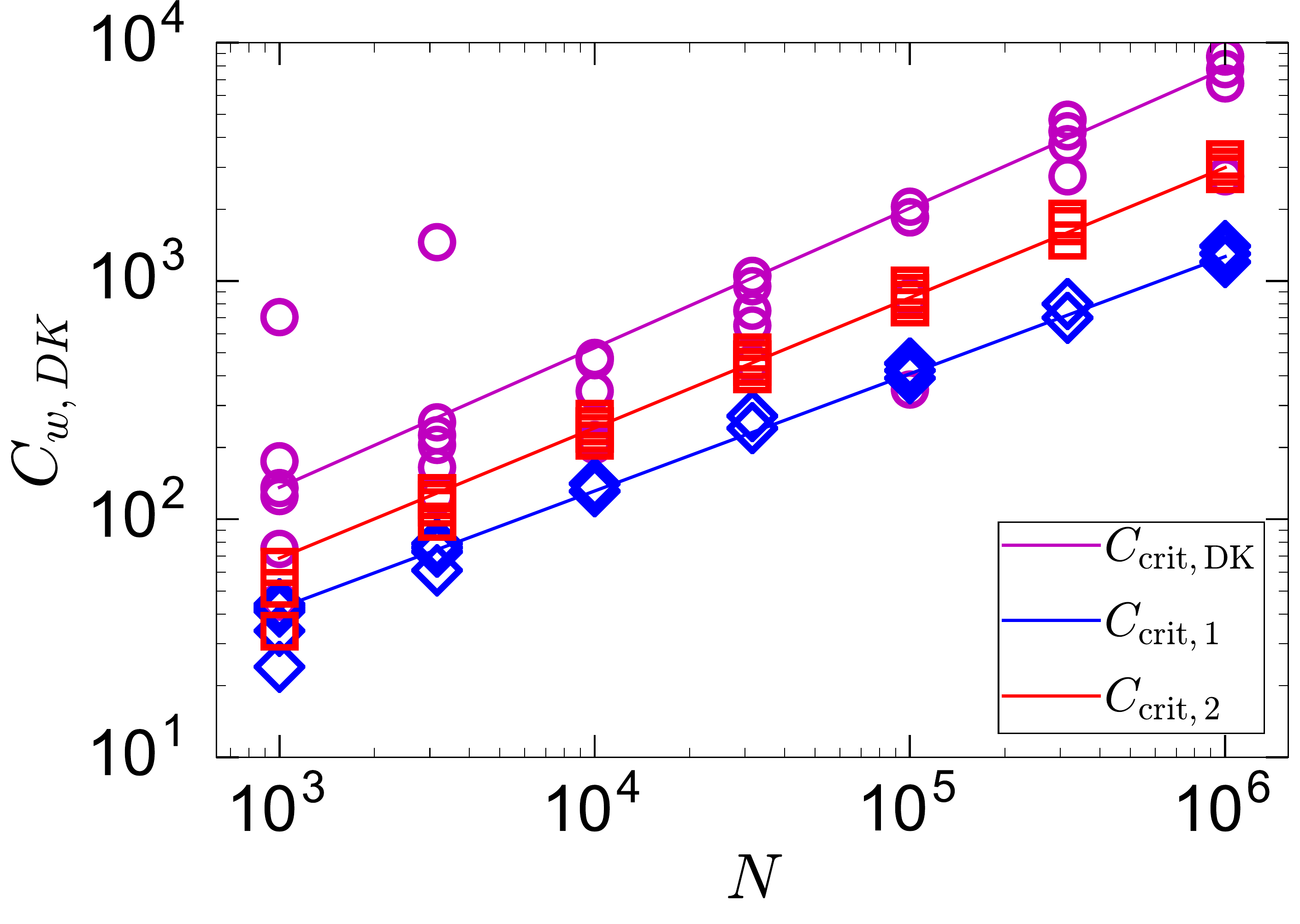}
	\caption{\label{fig:CwcritScaling}(Color online) Scaling of $C_\text{w,1}$ critical values versus $N$: $C_\text{crit,DK}$ (magenta circles), $C_\text{crit,1}$ (blue diamonds), and $C_\text{crit,2}$ (red squares). We plot critical values for $3.2\times10^{-4}\le \epsilon\le1.0\times 10^{-1}$, and fit those values to the model $a\times N^b$, where $b=0.59\pm0.03$ for simulations, $b=0.49\pm0.01$ for the one-step failure cascade theory, and $b=0.55\pm0.01$ for the two-step failure cascade theory.
}
\end{figure}

\begin{figure}[tb]
	\centering
  \includegraphics[width=1\columnwidth]{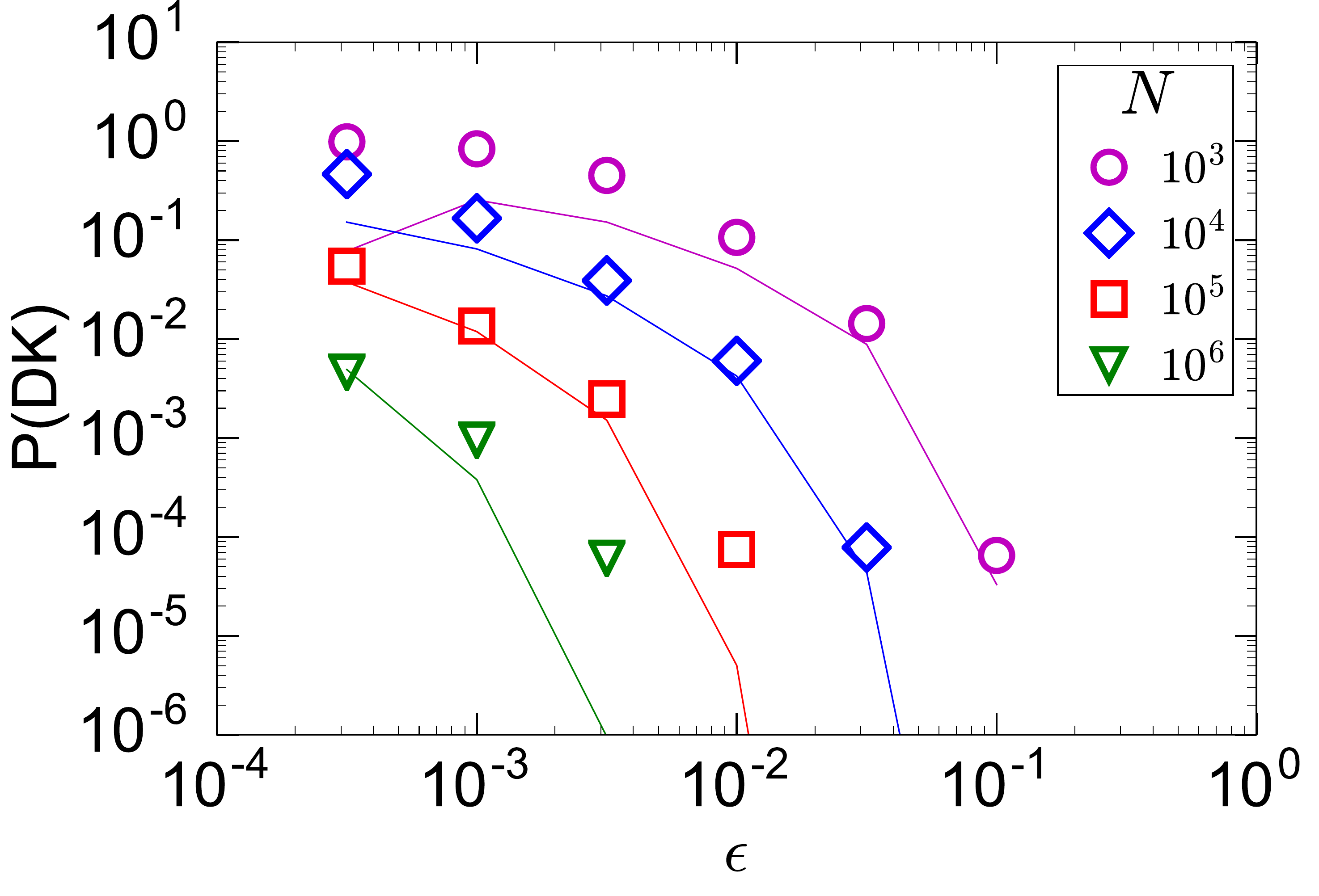}
	\caption{\label{fig:PDKVsEpsilon}(Color online) The probability of DKs as a function of epsilon. Regardless of N, we find close agreement between simulations-based probabilities of $P(DK)$ (open symbols) and theory (Eq.~\eqref{eq:pdkepsilon}, solid lines), especially for moderate values of $\epsilon$.
}
\end{figure}

Finally we can compare the two-step failure cascade theory to the one-step failure cascade theory and to simulations of DKs. The first thing we notice is that, as expected, the agreement with simulations of DKs is not perfect, especially for large $N$ (cf.~Fig.~\ref{fig:PCw2Cw3VsCw1}(b)), because the theory is still a necessary but not sufficient condition for DKs, although it is a significant improvement over the one-step failure cascade (cf.~Fig.~\ref{fig:PCw2Cw3VsCw1}(a)). We also notice that the critical values of $C_\text{w,1}$ for the probability of the one-step cascade to be 1/2, $C_\text{crit,1}$, the critical values of $C_\text{w,1}$ for the two-step cascade probability to be 1/2, $C_\text{crit,2}$, and the critical values of $C_\text{w,1}$ for the probability of a DK to be 1/2, $C_\text{crit,DK}$, all scale differently with system size. As Eq.~\eqref{eq:Cw1} shows, in the one-step failure cascade theory, $C_\text{crit,1}\sim N^b$ where $b=0.5$, while we find that, for the two-step failure cascade theory, $b=0.55\pm0.01$, and for simulations of DKs, $b=0.59\pm0.03$ (cf.~Fig.~\ref{fig:CwcritScaling}). To see more explicitly how two-step failure cascade theory and one-step failure cascade theory differ in their scaling, we can look at Fig.~\ref{fig:PDKVsEpsilon}, where it is clear that the two-step failure cascade theory reaches probability $1/2$ at larger and larger values of $C_\text{w,1}$ as $N$ increases, compared to the one-step failure cascade theory.

\subsection{III: How the Probability of Dragon Kings Varies With N and Epsilon}
\begin{figure}[tb]
	\centering
  \includegraphics[width=1\columnwidth]{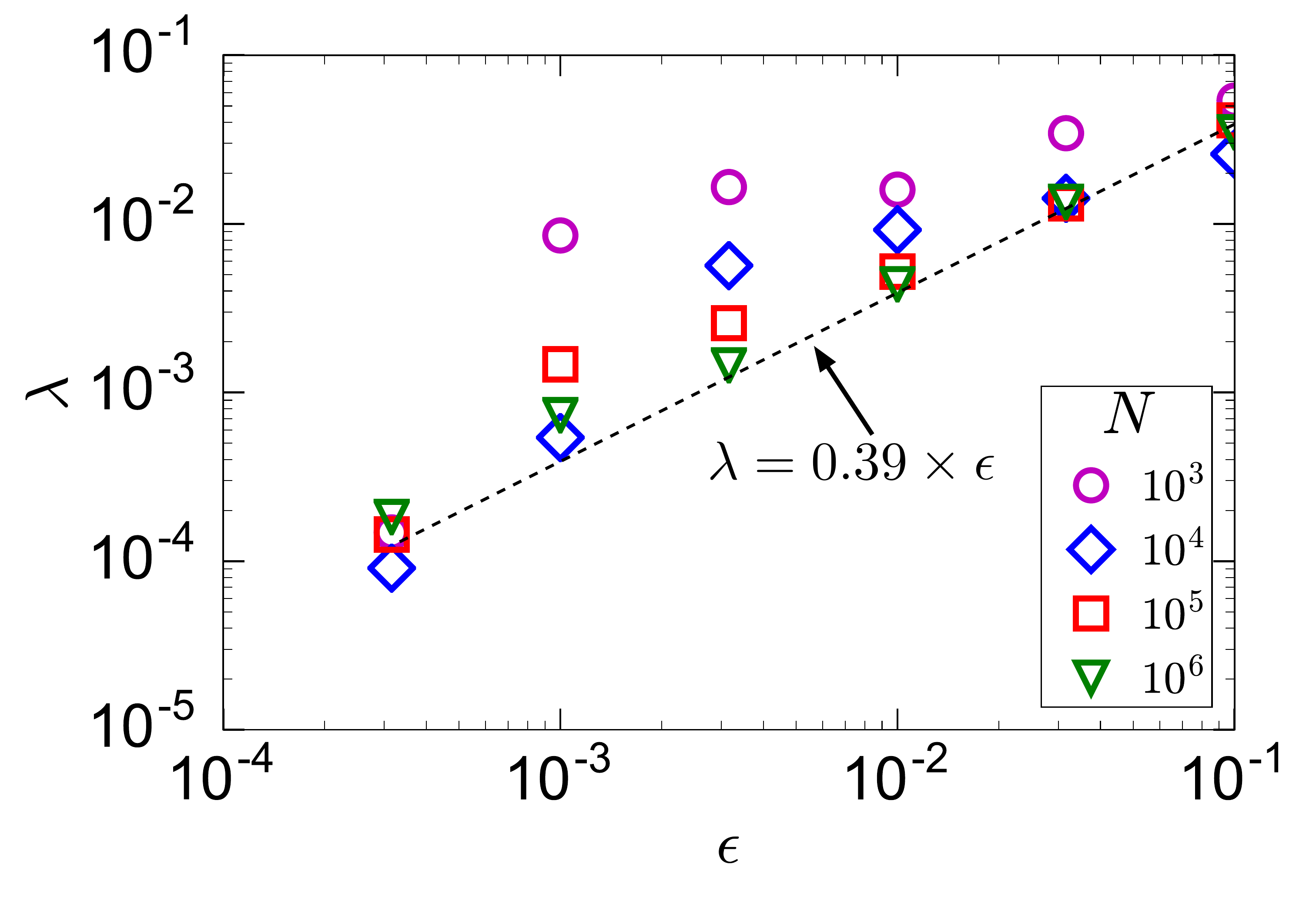}
  \caption{\label{fig:lambdaVsepsilon}The exponential cut-off parameter, $\lambda$, for the weak-node cluster distribution, $P(C_\text{w})$, versus $\epsilon$ for the CC model (cf.~Fig.~\ref{fig:weak_cluster} for $P(C_\text{w})$). The best-fit relation between $\lambda$ and $\epsilon$ is $\lambda=0.39\times \epsilon$ for $N=10^5$ and $10^6$.}
\end{figure}

When we plot the weak-node cluster size distribution, $P(C_\text{w})$, based on maximum likelihood estimates we find that 
\begin{equation}
P(C_\text{w})\sim C_\text{w}^{-\eta} e^{\lambda C_\text{w}},
\end{equation}
 where $\eta=2.26\pm0.03$ and we find that $\lambda=(0.39 \pm 0.01)\times \epsilon$  when $N=10^5$ and $10^6$ (cf.~Fig.~\ref{fig:lambdaVsepsilon}), and $\eta$ is measured with $\epsilon=3.2\times 10^{-4}$. This is consistent with our hypothesis that, in the duel limit that $N\rightarrow \infty$ and $\epsilon\rightarrow 0$, the model approaches a self-organized critical state, where the distribution of $C_\text{w,1}$ becomes a power law. 

The probability for the size of the first weak cluster that fails, $P(C_\text{w,1})$, is equivalent to picking a node in a cluster of size $C_\text{w}$, leading to
\begin{equation}
P(C_\text{w,1})= \frac{C_\text{w} P(C_\text{w})}{\langle C_\text{w} \rangle},
\end{equation}
or
\begin{equation}
P(C_\text{w,1})=  \frac{C_\text{w}^{1-\eta} e^{\lambda C_\text{w}}}{E_{\eta-1}(\lambda)},
\end{equation}
where
\begin{equation}
E_{n}(z) = \int_1^N \frac{e^{-z t}}{t^n} dt.
\end{equation}

Furthermore, we can approximate $P(\text{DK}|C_\text{w,1})$ as a step function,
\begin{equation}
P(\text{DK}|C_\text{w,1})\approx H(C_\text{w,1}-C_\text{crit,DK}).
\end{equation}
With these assumptions, we can approximate $P(\text{DK}|N,\epsilon)$ as
\begin{equation}
P(\text{DK}|N,\epsilon) = \int_{C_\text{crit,DK}}^N P(C_\text{w,1}) dC_\text{w,1}.
\end{equation}

If $\epsilon > 0$, then we can further approximate the integral limit as $N\rightarrow \infty$. This implies that 
\begin{equation}
\label{eq:pdkepsilon}
P(\text{DK}|N,\epsilon) =\begin{cases}
\lambda^{\eta-1} \frac{\Gamma(1-\eta,C_\text{crit,DK} \lambda)}{E_{\eta-1}(\lambda)}& \epsilon > 0\\
\frac{N^{\eta } \left(C_\text{crit,DK}^{2-\eta }-N^{2-\eta }\right)}{N^{\eta }-N^2} & \epsilon \rightarrow 0.\\
\end{cases}
\end{equation}

We notice two interesting findings. First, we see that $P(\text{DK})$ varies non-linearly with $\epsilon$, meaning a slight increase in repair frequency can dramatically reduce the number of system-wide failures. Agreement is strongest when $\epsilon$ is moderate, possibly because, when $\epsilon$ is too small for small $N$, the largest values of $C_\text{w}$ are $O(N)$, therefore the model under-estimates how many DKs are possible. When $\epsilon$ is too large, however, we appear to underestimate $P(\text{DK})$ again, potentially because the Heaviside approximation breaks down for small probabilities. We nonetheless find good overall agreement with our simulations.

Another interesting implication of our model is that for $\epsilon \rightarrow 0$ and $N\rightarrow \infty$, 
\begin{equation}
P(\text{DK}|N,0)\sim C_\text{crit,DK}^{\eta-2}-O(N^{2-\eta}).
\end{equation}
Through simulations, we find that $C_\text{crit,DK}\sim N^{0.59\pm 0.03}$, therefore

\begin{equation}
P(\text{DK}|N,0)\sim N^{-\beta} - O(N^{2-\eta}),
\end{equation}
where $\beta = 0.59 (2-\eta) = 0.15\pm0.02$. In other words, we find that DKs do not exist in the thermodynamic limit, but there are unusually large finite size effects. For example, this scaling implies that $P(\text{DK})$ decreases by only a factor of 10 when the system increases in size by ten million. For any real system DKs will exist and an engineer would be weary to ignore these finite size effects.

\begin{figure}[t]
	\centering
  \includegraphics[width=1\columnwidth]{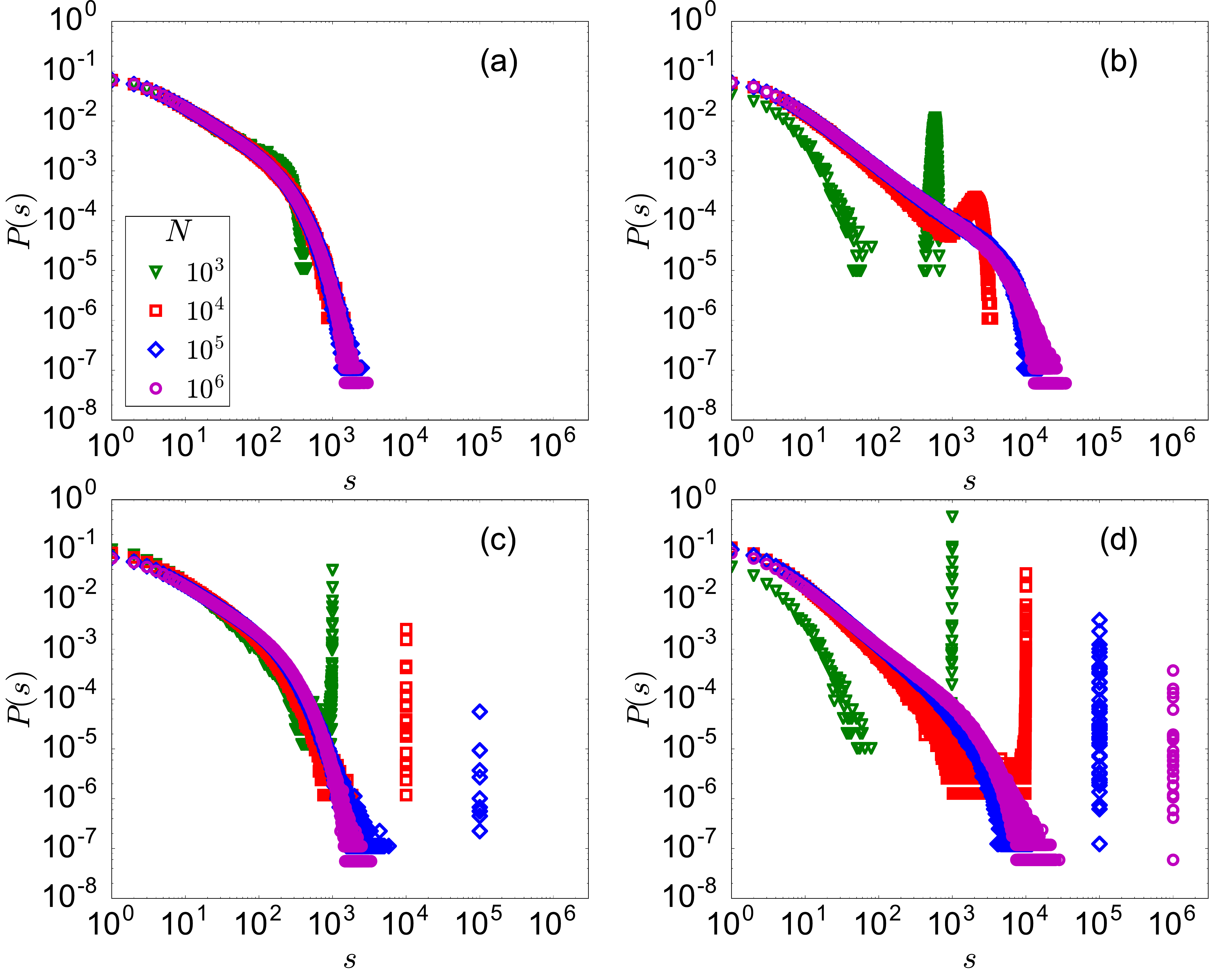}
	\caption{\label{fig:failure_eN}(Color online) Failure size distribution for networks with different $N$ and different reinforcement probability $\epsilon$. (a) IN model, $\epsilon=1.0\times10^{-2}$; (b) IN model, $\epsilon=1.0\times10^{-3}$; (c) CC model, $\epsilon=1.0\times10^{-2}$; (d) CC model, $\epsilon=1.0\times10^{-3}$;}
\end{figure}

\begin{figure}[t]
	\centering
  \includegraphics[width=1\columnwidth]{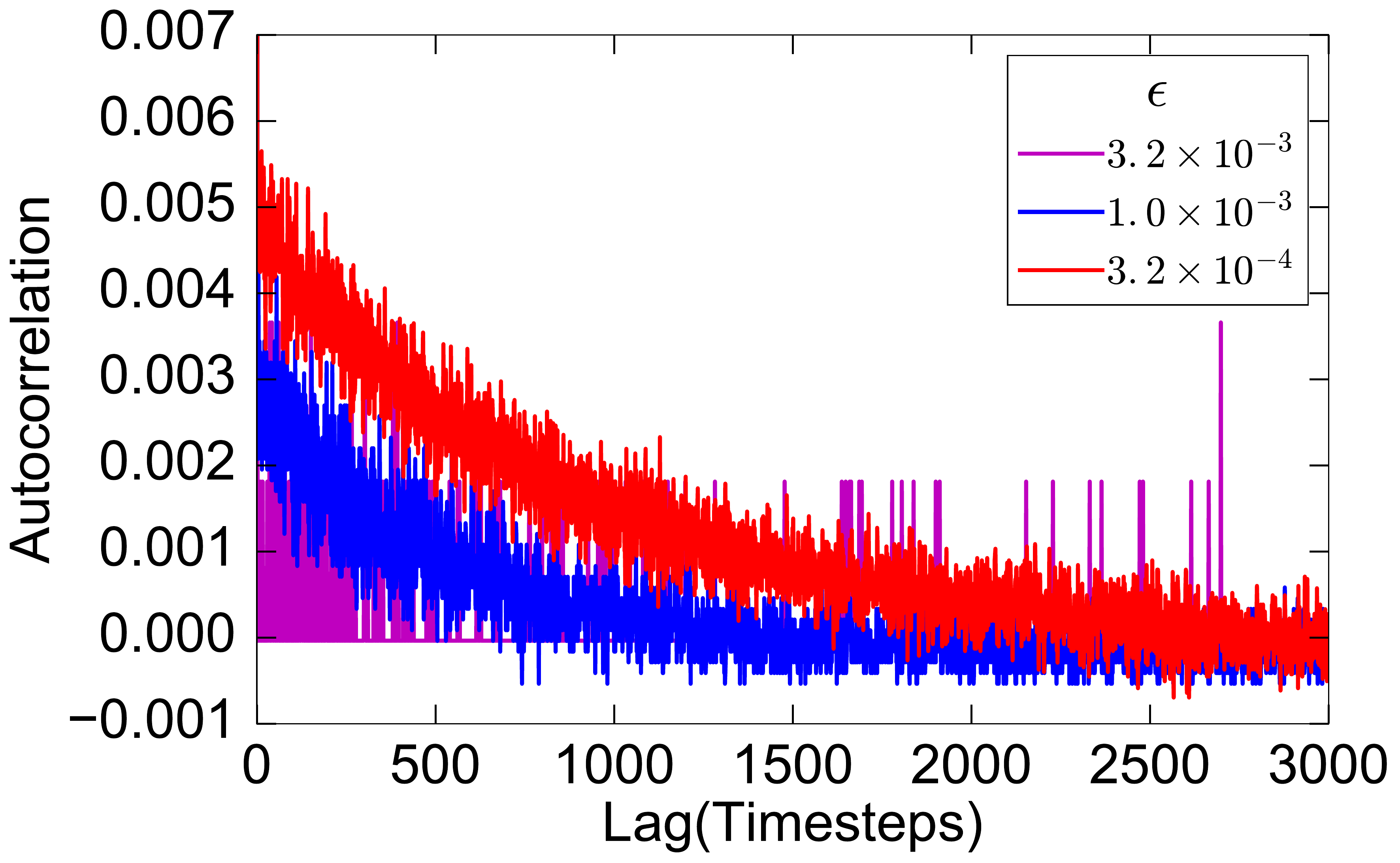}
  \caption{\label{fig:autocorrelation}DK autocorrelation versus lag time. We find that, regardless of the value of $\epsilon$ and regardless of the lag time, the autocorrelation of DKs over $1.5\times 10^7$ timesteps is very low, especially as $\epsilon$ increases. We do not plot autocorrelations for $\epsilon \ge 1.0\times 10^{-2}$ because there are few DKs for these values of $\epsilon$ in the time frame studied. The network size is $N=10^6$.}
\end{figure}

\begin{figure}[t]
	\centering
  \includegraphics[width=0.95\columnwidth]{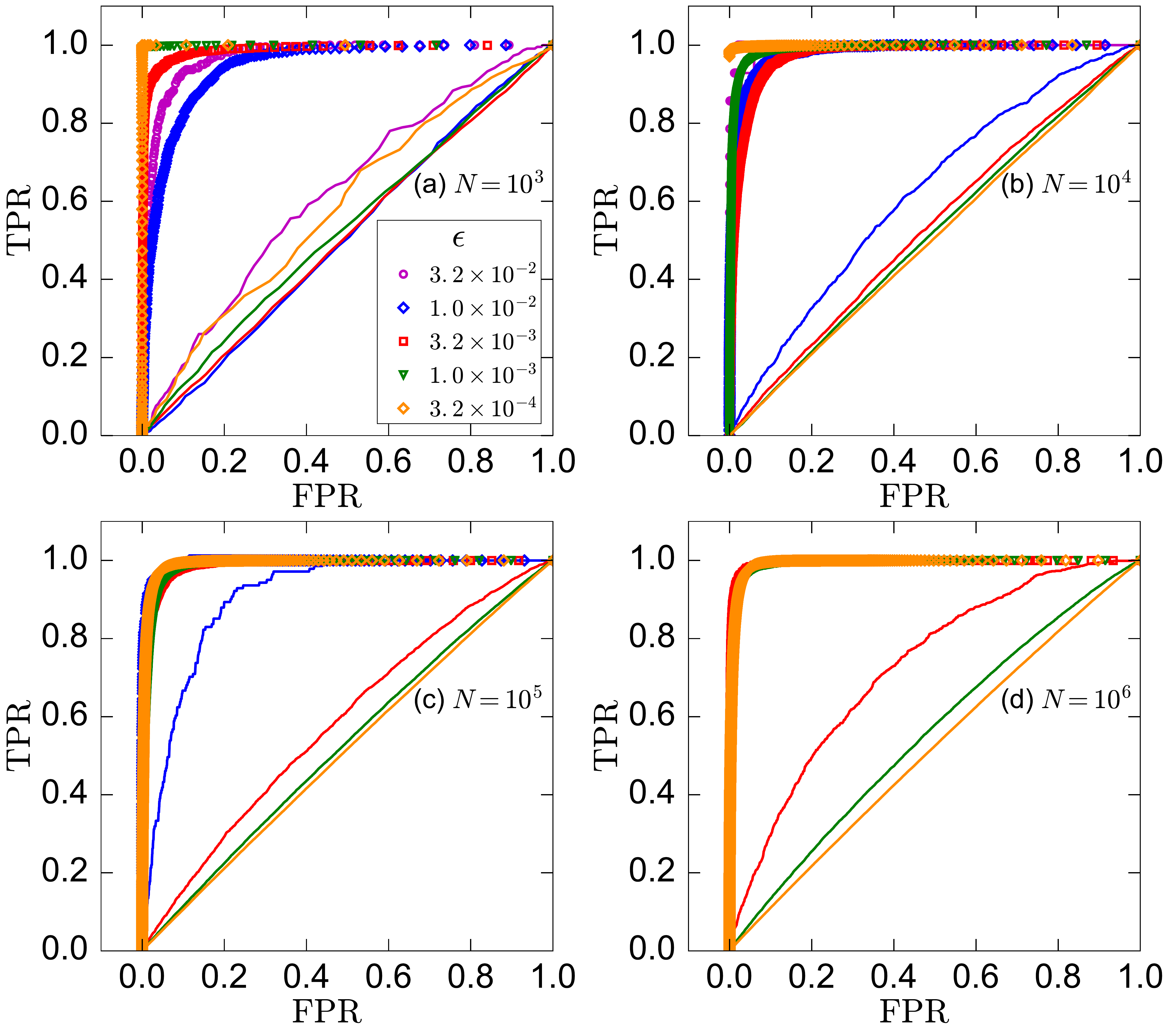}
	\caption{\label{fig:roc}(Color online) ROC curves for different $N$ and different $\epsilon$ using two different predictors. The solid lines are the results predicted by the fraction of weak nodes, and the open markers represent the results predicted by the size of the first weak-node cluster.
}
\end{figure}

\begin{figure}[t]
	\centering
  \includegraphics[width=1\columnwidth]{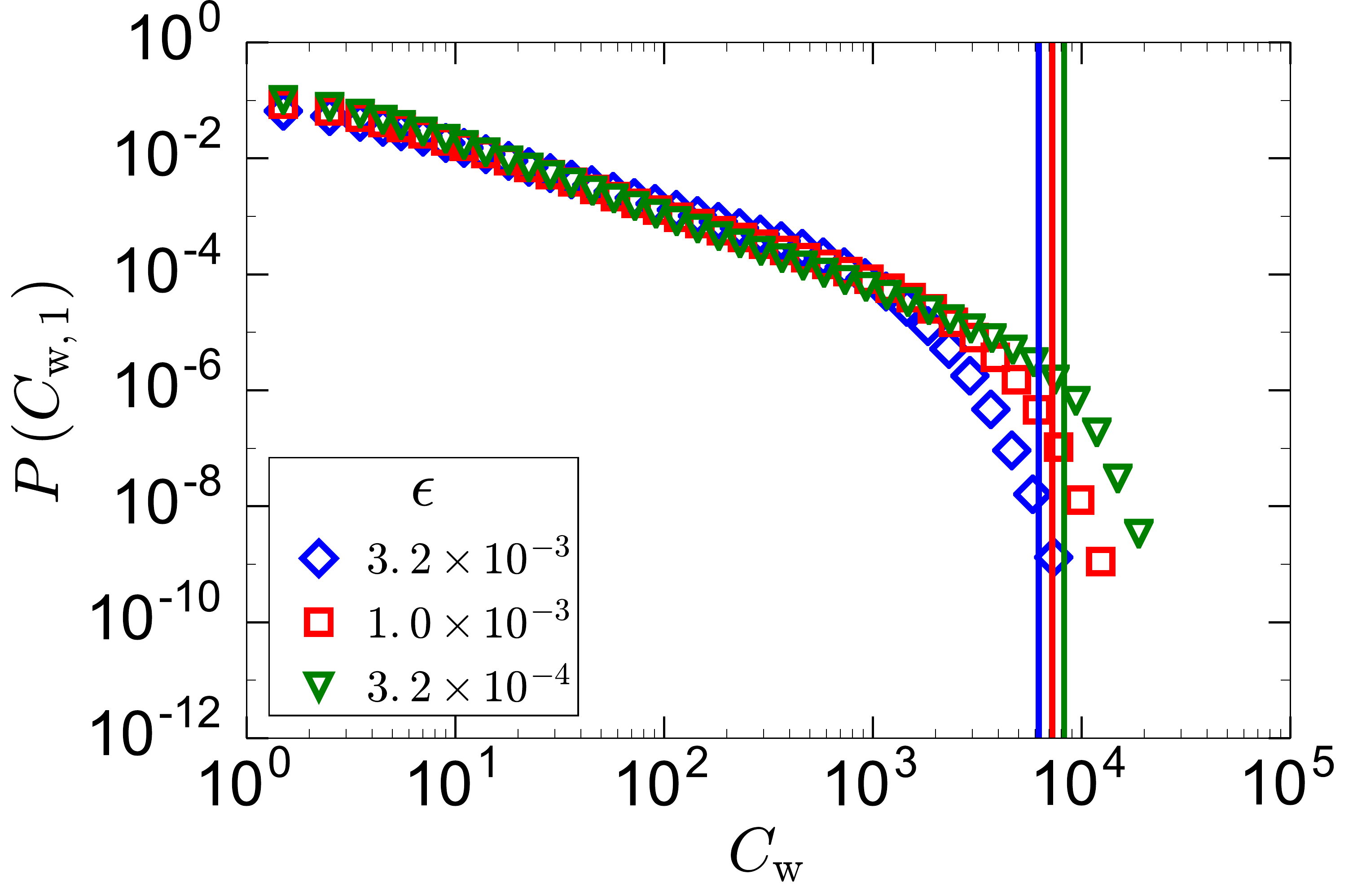}
	\caption{\label{fig:cw1}(Color online) The probability the first weak-node cluster that fails is of size $C_\text{w}$, for $N=10^6$ over $15\times N$ timesteps and $1$ network realization. The vertical lines show the value of $C_\text{crit,DK}$.
}
\end{figure}

\begin{figure}[t]
	\centering
  \includegraphics[width=1\columnwidth]{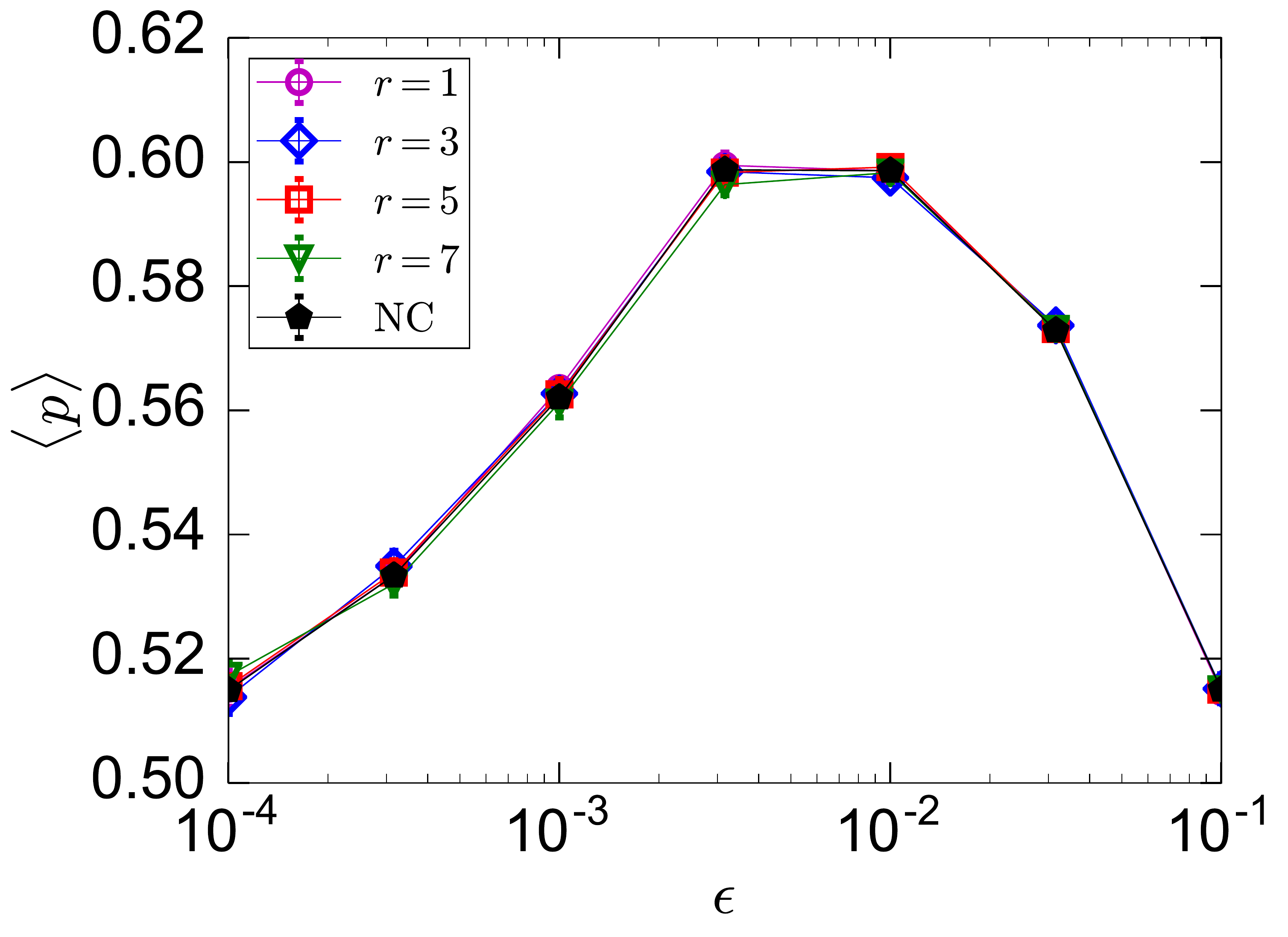}
	\caption{
    \label{fig:average_control}(Color online) The fraction of weak nodes averaged over $T$ timesteps with and without control for $N=10^6$ and 10 network realizations. For the uncontrolled system, $T=5\times N$, while for the controlled system $T=10\times N$.
}
\end{figure}

\subsection{\label{sec:II}IV: Failure Distribution For Finite System Size}

Figure 1 in the main text demonstrates that DKs in the CC model occur for various values of $\epsilon$, and our analytic arguments show DKs occur for all values of $\epsilon < 1$ and $N < \infty$. We find a superficially similar failure size distribution, with a bump of size $O(N)$, for the IN model (see Figs.\ \ref{fig:failure_eN}(a) \& \ref{fig:failure_eN}(b)) but only when $N$ and $\epsilon$ are sufficiently small, e.g., $N=10^4$ for $\epsilon\le 10^{-3}$. This is not due to cascading failure, however, because in the IN model only a single weak-node cluster fails. This bump exists because there are weak-node clusters that are $O(N)$, meaning we are in a parameter regime where there are super-critical percolating clusters.

This contrasts with the CC model, where we see over $99.9\%$ of nodes fail almost independent of the values of $\epsilon$ and $N$ (see Figs.\ \ref{fig:failure_eN}(c) \& \ref{fig:failure_eN}(d)). The difference is due to cascading failures in the CC model, where the moment a cluster greater than a critical size fails, strong nodes begin to fail, which triggers more weak-node clusters to fail, etc., until almost all nodes fail (a DK event). Because the critical weak-node cluster size increases sub-linearly with $N$, and because weak-node clusters can be any size less than $N$, there is always a chance for weak-node clusters larger than the critical size to fail, triggering a DK event.

It is an open question in the field of DK theory how to further classify events such as the bump in the IN model, which has a heavier-than-power-law probability, yet shares the same underlying mechanism as small events (i.e., the mechanism being the failure of a single cluster of weak nodes in the IN model.)

\subsection{V: Receiver operating characteristic curves of all prediction methods}

In the section, \emph{Predicting Dragon Kings}, in the main text, we use the area under the receiver operating characteristic curve (AUC) to compare the accuracy of two predictors: the fraction of weak nodes, $p$, and the size of the first weak-node cluster, $C_\text{w,1}$. We briefly mention that DKs are not highly correlated in time (Fig. \ref{fig:autocorrelation}), therefore it is non-trivial exercise to find when DKs occur. Examples of ROC curves are illustrated in Fig.~\ref{fig:roc}. The ROC curve is created by plotting the true positive rate (TPR) against the false positive rate (FPR) as the discrimination threshold is varied \cite{Brown2006}. Here, the TPR indicates the probability of DKs being correctly predicted. The FPR is the probability of DKs being wrongly predicted. We conclude that the second predictor, $C_\text{w,1}$, is close to being an optimal predictor. Moreover, DKs are predicted by $C_\text{w,1}$ with high accuracy, because DKs are unlikely to occur for $C_\text{w,1}<C_\text{crit,DK}$, and due to the power-law distribution of $C_\text{w,1}$ as shown in Fig.~\ref{fig:cw1}, most initial failures lead to small cascades. For example, when $\epsilon=1.0\times10^{-3}$, the probability $P(C_\text{w,1}<C_\text{crit,DK})$ is $99.986\%$.

\subsection{VI: Average fraction of weak nodes with or without control}
In our work, we analyze a control strategy for DKs. Our objective is to find a strategy that keeps the amount of weak nodes constant on average compared to the CC model, but reduces the frequency of DKs. We randomly pick $r$ weak nodes and consider the size of the weak-node clusters to which they belong. The largest of these weak-node clusters is selected and with probability $1-p(t)$ and a random node in that cluster is reinforced. In Fig.~\ref{fig:average_control}, we demonstrate that varying $r$ does not significantly change the average fraction of weak nodes, even though it significantly affects the prevalence of DKs, as shown in the main text.

%\bibliographystyle{apsrev4-1}
%\bibliography{DK.bib}
%

\end{document}